\documentclass[psfig]{aa}
\input psfig.tex

\begin{document}                                                                

\onecolumn

%-------------------- own definitions -----------------------                   
% ------------------------------------------------                              
%   \thesaurus{06         % A&A Section 6: Form. struct. and evolut. of stars    
%              (03.11.1)}  % Cosmogony,                                         

   \title{Cosmic Mass Functions from Gaussian Stochastic Diffusion Processes}

   \titlerunning{Cosmic Mass Functions from Gaussian Stochastic Diffusion Processes}

   \author{P.\,Schuecker, H.\,B\"ohringer, K.\,Arzner  and T.H.\,Reiprich}
                                       
   \authorrunning{Schuecker et al.}

   \offprints{Peter Schuecker\\ peters@mpe.mpg.de} 

   \institute{Max-Planck-Institut f\"ur extraterrestrische Physik, 
                   D-85741 Garching, Germany}

   \date{Received 30.06.00; accepted 23.02.2001}

   \markboth {Cosmic Mass  Functions from Gaussian Stochastic Diffusion Processes}{}

\abstract{Gaussian stochastic diffusion processes are used to derive
cosmic mass functions. To get analytic relations previous studies
exploited the sharp $k$-space filter assumption yielding zero drift
terms in the corresponding Fokker-Planck (Kolmogorov's forward)
equation and thus simplifying analytic treatments significantly
(excursion set formalism). In the present paper methods are described
to derive for given diffusion processes and Gaussian random fields the
corresponding mass and filter functions by solving the Kolmogorov's
forward and backward equations including nonzero drift terms. This
formalism can also be used in cases with non-sharp $k$-space filters
and for diffusion processes exhibiting correlations between different
mass scales.   }

    \maketitle

\section{Introduction}\label{INTRO}

The Press-Schechter (1974) formula of the cosmic mass function has
found many important applications to basic problems in modern
cosmology (e.g., Narayan \& White 1988, Cole \& Kaiser 1988, 1989,
Henry \& Arnaud 1991, White \& Frenk 1991, White et al. 1993, Eke et
al. 1996, Mo et al. 1996, Mo \& White 1996, Matarrese et al. 1997,
Mathiesen \& Evrard 1998, Borgani et al. 1999). Although the approach
appears questionable when the theoretical masses are compared with the
group finder masses in detailed N-body simulations (e.g., Bond et
al. 1991, White 1996) the overall statistical agreement is found to be
sufficiently good (Efstathiou et al. 1988, Lacey \& Cole 1994, but see
also Sect.\,\ref{DISCUSS}). Improved versions relax the various
assumptions of the original derivation (Lucchin \& Matarrese 1988,
Bond et al. 1991, Lilje 1992, Cavaliere \& Menci 1994, Jedamzik 1995,
Monaco 1995, Kitayama \& Suto 1996, Yano, Nagashima \& Gouda 1996,
Valageas \& Schaeffer 1997, Chiu et al. 1998, Lee \& Shandarin 1998).

Apart from providing a convenient fitting equation, the excursion set
formalism in the form presented in Bond et al. (1991) serves as the
basis for several fundamental theoretical developments aiming for a
full treatment of structure formation in a hierarchically clustering
universe. Extended theories for hierarchical clustering (Bower 1991),
biasing schemes (Mo \& White 1996), and merging histories (Lacey \&
Cole 1993, Kauffmann \& White 1993) are developed or can also be
derived in the end by assuming the validity of the excursion set
arguments introduced by Bond et al. to derive the Press-Schechter
formula with the correct normalization and several related conditional
probability densities.

For the derivation of the Press-Schechter function the excursion set
formalism assumes that at a fixed spatial location the change of the
density contrast successively smoothed with a decreasing filter scale,
$R$, can be described as a sample path of an abstract diffusion
process (Bond et al.  1991, see also Peacock \& Heavens 1990). The
counting of subclumps (clouds-in-clouds) is avoided by computing the
rates at which the sample paths meet at their largest mass extent a
scale-independent critical density contrast barrier, $f_{\rm c}$. For
this process Bond et al.  derived a general Fokker-Planck
(Kolmogorov's forward) equation consisting of a drift and of a
diffusion term. The application of the sharp $k$-space filter
condition yields a zero drift term so that the resulting much simpler
Fokker-Planck equation can be solved easily and leads to the
Press-Schechter mass function with the correct normalization.

Unfortunately, analytic results are obtained with the excursion set
formalism only within the theoretical framework of Markovian processes
(see below) and for the sharp $k$-space filter with a mass assignment
scheme which clearly contradicts numerical experiments on the
halo-by-halo basis. It is the basic aim of the present investigation
to extend the excursion set formalism to non-sharp $k$-space filters
but without leaving the framework of Markovian random processes and
thus preserving the simplicity of the original approach.  In the
present context, Markov processes are assumed to be driven by
deterministic and stochastic `forces', and have the simplifying
property that when the process states, in form of amplitudes of
filtered density contrasts, are known on a given mass scale and
larger, the process states on smaller scales depend only on the actual
scale and are {\it independent} of the larger scales -- it seems
suprising that under this Markov condition, e.g., useful biasing
schemes can be derived, but see below.

The present paper discusses both the Kolmogorov's forward and backward
equation including nonvanishing drift terms. Analytic solutions are
presented for several interesting cases offering the possibility to
derive cosmic mass functions independent of the sharp $k$-space filter
assumption. It will be shown that certain Gaussian stochastic
diffusion processes lead to more realistic filter profiles and to
correlations between different mass resolution scales which are
expected to have important implications on various progenitor
statistics and biasing schemes.

Section\,\ref{EXCSET} gives an informal presentation of the basic
concepts of the excursion set formalism. Section\,\ref{GENERAL}
summarizes the formal properties of the relevant diffusion processes
and their sample paths. The connection between diffusion processes and
mass functions is outlined in Sect.\,\ref{DPMF}. The relation between
the basic pseudo time (resolution) variable and the filter radius is
discussed in Sect.\,\ref{SHARP}. In Sects.\,\ref{MASS_WIEN} and
\ref{MASS_OU} the methods are applied to order-0 (Wiener) diffusion
processes, characterized by drift coefficients which are independent
of mass and filtered density contrast, and to order-1
(Ornstein-Uhlenbeck) diffusion processes, characterized by drift
coefficients which are also independent of mass but which scale
linearly with filtered density contrast. For both processes the
corresponding mass and filter functions are derived. The results are
summarized and discussed in Sect.\,\ref{DISCUSS}. Appendix A gives an
overview of the basic properties of order-1 diffusion processes.

Forthcoming papers will analyse the new mass assignment scheme, and
will compare the mass functions presented here and others deduced from
more complicated stochastic diffusion processes with detailed N-body
simulations and with observed mass and luminosity functions of rich
clusters of galaxies obtained with, e.g., the REFLEX cluster sample
(B\"ohringer et al. 1998, Guzzo et al. 1999), and the HIFLUGCS cluster
sample (T.H. Reiprich \& H. B\"ohringer, in preparation).

\section{Excursion set formalism}\label{EXCSET}

The differential mass function, $n(M)$, gives the number of virialized
objects of mass $M$, per unit mass and (comoving) volume element.  We
want to derive $n(M)$ using the field of density contrasts,
$f(\vec{r})$, at the comoving coordinates, $\vec{r}$. Here, $f$ is
defined by the ratio $f(\vec{r})=\rho(\vec{r})/\bar{\rho}-1$, where
$\rho(\vec{r})$ is the comoving mass density at $\vec{r}$, and
$\bar{\rho}$ the mean comoving mass density. The following notation is
simplified by neglecting any evolution of $f$ with cosmic
time. However the final equations can be transformed to any epoch by
multiplication with linear growth factors. For the derivation of
$n(M)$ it is assumed that the location where an virialized object of
mass $M$ will be formed corresponds to the location where
$f(\vec{r})$, filtered with the radius $R(M)$ to give
$\tilde{f}(\vec{r},R)$, has reached a universal critical threshold
value, $f_{\rm c}$. The density field is divided into infinitesimal
small volume elements. A sequence of filtered three-dimensional
density fields is constructed with gradually decreasing filter
radii. The excursion set formalism identifies the cumulative mass
fraction $\propto n(>M)$ with the fraction of volume elements in which
the density contrast lies above $f_{\rm c}$ when smoothed with any
filter of radius greater than or equal $R(M)$.

At a fixed spatial location, $\vec{r}$, the sequence of filtered
density contrasts, $\{\tilde{f}(\vec{r},R)|R=\infty\ldots 0\}$, can be
regarded as a one-dimensional trajectory. The statistical description
of an ensemble of such sample paths gets especially simple when they
result from a {\it Gaussian} random process with independent
increments, i.e., from a Wiener process. For filtered density
contrasts this is the case when $f(\vec{r})$ is a Gaussian random
field uniquely described by the fluctuation power spectrum, $P(k)$, as
a function of comoving wavenumber $k$, and when the top-hat filter in
Fourier space, i.e., the sharp $k$-space filter (see Eq. \ref{SHARPK}
below) is used:

The power spectrum of the density field is defined via the Fourier
transform of the unfiltered density contrasts, $f(\vec{k})={\rm
FT}[f(\vec{r})]$, by the expectation value ${\rm
E}[f(\vec{k})\,f^*(\vec{k}')]=(2\pi)^3P(k)\delta_{\rm
D}(\vec{k}-\vec{k}')$, where $\delta_{\rm D}$ is the Dirac delta
distribution and $^*$ assigns the complex conjugate. In this case the
discrete change in $\tilde{f}(\vec{r},R)$ caused by the transition
$k_R\rightarrow k_R+\Delta k_R$ of the maximum comoving wavenumber,
$k_R$, which passes the filter is a Gaussian random variable (each
change of $\tilde{f}(\vec{r},R)$ when $R$ is changed comes from the
inclusion/exclusion of a new independent $k$ shell) with the variance
$\Delta\sigma^2=\sigma^2[1/(k_R+\Delta k_R)]-\sigma^2[1/k_R]$, where
\begin{equation}\label{VARI}
\sigma^2(R)\,=\,\frac{1}{(2\pi)^3}\,\int_0^\infty\,d^3k\,|W_R(k)|^2\,P(k)\,.
\end{equation}
The sharp $k$-space filter, 
\begin{equation}\label{SHARPK}
W_R(k)\,=\,\left\{ \begin{array}{r@{\quad:\quad}l}
1 & k\leq k_R=1/R \\ 0 & {\rm else} \end{array} \right.\,, 
\end{equation}
and Eq. (\ref{VARI}) define (for spherically symmetric filters) a
general resolution variable,
\begin{equation}\label{RESOLVAR}
\Lambda(R)\,=\,\frac{1}{2\pi^2}\,\int_0^{\frac{1}{R}}\,dk\,k^2\,P(k)\,.
\end{equation}
For this choice, the $\tilde{f}(\Lambda)$ values perform some kind of
Brownian motion with $\Lambda(R)$ as a pseudo time variable. Notice
that for hierarchical clustering, $\Lambda$ as well as $\sigma^2$
increase with decreasing mass and filter scale. For $\Delta
k_R\rightarrow 0$ a trajectory corresponds to one realization of a
continuous limit of a simple random walk: the Wiener process. 

Hence, the behaviour of these trajectories are governed by a
simplified Fokker-Planck diffusion equation which gives the
probability density, $\Pi$, to find trajectories with
$\tilde{f}(\vec{r},R)$ at $\Lambda$. In this picture the differential
mass function can be obtained from the rate at which the random
trajectories meet for the first `time' the absorbing barrier $f_{\rm
c}$ at a given mass or $\Lambda[R(M)]$ scale. The final problem is to
get solutions of the Fokker-Planck equation under the boundary
conditions that all sample paths start at $\Lambda=0$ with
$\tilde{f}=0$ and that absorption of a sample path at $\Lambda$
corresponds to $\Pi(\tilde{f}=f_{\rm c},\Lambda)=0$ (see
Sect.\,\ref{MASS_WIEN} for a more detailed discussion of these
boundary conditions).

The excursion set formalism in the form presented above provides a
simple mean to deduce the Press-Schechter mass function. The major
reason for the simplicity of the prescription results from the
application of the sharp $k$-space filter. The choice of this filter
appears to be rather {\it ad hoc} and not thoroughly motivated. Its
strongly oscillating shape in configuration space does not seem to be
optimal to frame the region of primordial material that ultimately
collapses to form a virialized halo. Moreover, the absence of any
covariances between different mass scales is not expected for
realistic physical systems. In the following a more general (though
more formal) description of the evolution of $\Pi$ is given, by
considering a general diffusion process which offers the possibility
to study also non-sharp filter functions and the resulting mass
functions. It will be seen that certain diffusion processes lead to
nonzero covariances between different mass scales without leaving the
general framework of Markovian random processes. These processes are
thus expected to yield more realistic formation histories and biasing
schemes.

\section{Gaussian $\Lambda$-$\tilde{f}$ diffusion processes}\label{GENERAL}

A Gaussian random field of the mass density contrast, $f$, is filtered
on a continuous set of decreasing scales $R$, giving at a fixed
spatial location in configuration space a continuous sequence of
filtered density contrasts, $\tilde{f}$, at the scale $R$, providing a
measure of the spatial resolution, $\Lambda(R)$. For scale-invariant
power spectra, $P(k)\propto k^n$, the resolution variable has the form
$\Lambda\propto R^{-(n+3)}$ (Eq. \ref{RESOLVAR}). The diffusion
process is restricted to a continuous - $\Lambda$, continuous -
$\tilde{f}$ state space, where $\Lambda$ and $\tilde{f}$ assume the
r\^{o}les of the usual time and spatial coordinates, respectively. In
general it is expected that the sample paths of the diffusion process
start with $\tilde{f}=\tilde{f}_0=0$ at $\Lambda=0$. Formally, a
Gaussian $\Lambda$-$\tilde{f}$ diffusion process corresponds to a
Markov process for which the transition probabilities,
$\Pi(\tilde{f}',\Lambda+\Delta\Lambda,\tilde{f},\Lambda)$, with the
sharp value $\tilde{f}$ at $\Lambda$ and the random variable
$\tilde{f}'$ at $\Lambda+\Delta\Lambda$, satisfy a continuity
condition (similar to Eq. (a) in Arnold 1974, p.\,40), and
\begin{equation}\label{GEN1}
\lim_{\Delta\Lambda\to 0} \frac{1}{\Delta\Lambda}
\int_{|\tilde{f}'-\tilde{f}|\le\epsilon}\,
d\tilde{f}'\,(\tilde{f}-\tilde{f}')^m\,
\Pi(\tilde{f}',\Lambda+\Delta\Lambda,\tilde{f},\Lambda)
\,=\,A_m(\tilde{f},\Lambda)+{\rm O}(\epsilon)
\end{equation}
exists for $\epsilon>0$, $m=1,2$, and $A_m=0$ for $m\ge 3$. The $A_1$
coefficient describes the deterministic drift or the instantaneous
rate of change of the mean, and $A_2$ the stochastic diffusion or the
instantaneous rate of change of the squared fluctuations of the sample
paths giving the expectation, ${\rm
E}[\tilde{f}(\Lambda)-\tilde{f}(\Lambda')]^m\,=
\,A_m(\tilde{f},\Lambda)\Delta\Lambda+ \,O(\Delta\Lambda)$. Note that
the Markovian independence property introduces the simplifications
mentioned in White (1997) concerning the application of the extended
Press-Schechter theory to formation histories and biasing schemes:
Because the Markov property is used to derive extended
Press-Schechter, a say $10^{12}\,{\rm M_\odot}$ halo at redshift $z=1$
does not `know' whether it will be incorporated into a rich cluster or
remain in a void at $z=0$. The galaxy populations in protocluster and
protovoid halos will thus be the same. However, the present
investigation concentrates on a wider class of diffusion processes
than studied by the extended Press-Schechter approach, some of them
exhibiting nonzero correlations between different mass scales. The
expectation ${\rm E}[\tilde{f}(\Lambda)-\tilde{f}(\Lambda')]^m$ leads
with the inclusion of some additional heuristic arguments (see the
formal treatments in, e.g., Friedman 1975, Kloeden \& Platen 1995) to
the It\^o stochastic differential equation,
\begin{equation}\label{ITO}
d\tilde{f}(\Lambda)\,=\,-A_1(\tilde{f},\Lambda)d\Lambda\,+
\sqrt{A_2(\tilde{f},\Lambda)}\,dw(\Lambda)\,.
\end{equation}
The second term on the right-hand side is the fundamental or Wiener
stochastic process, i.e., a Gaussian process with the independent
infinitesimal increments,
$dw(\Lambda)\,=\,w(\Lambda+d\Lambda)-w(\Lambda)\,=\,K(\Lambda)d\Lambda$.
A prescription for the construction of $w(\Lambda)$ is given by
eqs.\,(\ref{PATHS}) and (\ref{BROWN}). The stochastic `forces',
$K(\Lambda)$, are Gaussian with {\it infinite} variance and no
correlations between different $\Lambda$ shells. They drive the
$\tilde{f}$-$\Lambda$ diffusion as illustrated by the Langevin-type
equation, $d\tilde{f}/d\Lambda\,=\,-A_1(\tilde{f},\Lambda) +
\,\sqrt{A_2(\tilde{f},\Lambda)}\,K(\Lambda)$. The increments,
$d\tilde{f}$, are of the order of the standard deviations of the
fundamental process, that is, $d\tilde{f}/d\Lambda\approx
\sqrt{\Delta\Lambda}/\Delta\Lambda$, which tends to infinity as
$\Delta\Lambda\rightarrow 0$ (see also Wiener 1930, Doob 1942). The
It\^o formalism is used to handle the resulting almost surely
continuous but almost surely nowhere differentiable sample paths in a
mathematically well-defined way. An alternative would be the
application of the Stratonovich formalism, but the results derived
here do not depend on the specific choice.

Closely related to the stochastic diffusion Eq. (\ref{ITO}) are
Kolmogorov's forward (Fokker-Planck) equation,
\begin{equation}\label{FP2}
\frac{\partial\,\Pi(\tilde{f}_0,\Lambda_0,\tilde{f},\Lambda)}
{\partial\Lambda}\,=\,{\rm\cal{L}}(\tilde{f},\Lambda)\,
\Pi(\tilde{f}_0,\Lambda_0,\tilde{f},\Lambda)\,,\quad\quad
{\rm\cal{L}}(\tilde{f},\Lambda)\,=\,\frac{\partial}{\partial\tilde{f}}
A_1(\tilde{f},\Lambda)+\frac{1}{2}\frac{\partial^2}{\partial\tilde{f}^2}
A_2(\tilde{f},\Lambda)\,,
\end{equation}
with the forward variables, $\tilde{f}$ and $\Lambda$, and
Kolmogorov's backward equation, 
\begin{equation}\label{FP2a}
\frac{\partial\,\Pi(\tilde{f}_0,\Lambda_0,\tilde{f},\Lambda)}
{\partial\Lambda_0}\,=\,-{\rm\cal{L}}^\dagger(\tilde{f}_0,\Lambda_0)\,
\Pi(\tilde{f}_0,\Lambda_0,\tilde{f},\Lambda)\,,\quad\quad
{\rm\cal{L}}^\dagger(\tilde{f}_0,\Lambda_0)\,=\,
-A_1(\tilde{f}_0,\Lambda_0)\frac{\partial}{\partial\tilde{f}_0}
+\frac{1}{2}A_2(\tilde{f}_0,\Lambda_0)\frac{\partial^2}{\partial\tilde{f}_0^2}\,,
\end{equation}
with the backward variables, $\tilde{f}_0$ and $\Lambda_0$.  Here,
${\rm\cal{L}}^\dagger$ is defined as the formal adjoint of the
elliptical operator ${\rm\cal{L}}$ (e.g., Kloeden \& Platen 1995,
Sect.\,2.4, Arnold 1974, Sect.\,2.6). In Eq. (\ref{FP2}) the
backward variables $\tilde{f}_0$ and $\Lambda_0$, and in (\ref{FP2a})
the forward variables $\tilde{f}$ and $\Lambda$ are essentially
constant and enter only through boundary conditions. The sign of $A_1$
is chosen to be consistent with (\ref{ITO}). The It\^o formula (see
Eq. \ref{ITOFORM} in Appendix A) can be used to verify the
equivalence of the Fokker-Planck equation and (\ref{ITO}). For
$\tilde{f}$-$\Lambda$ diffusion processes Bond et al.  (1991) showed
that Eq. (\ref{FP2}) can be derived via the Chapman-Kolmogorov
equation in the standard way. For these diffusion processes it is also
straightforward to show the validity of Kolmogorov's backward
equation.  Hence, (\ref{ITO}) can be used also for more general
$\tilde{f}$-$\Lambda$ diffusion processes.

\begin{figure}
\vspace{-2.0cm}
\centerline{\hspace{-9.0cm}
\psfig{figure=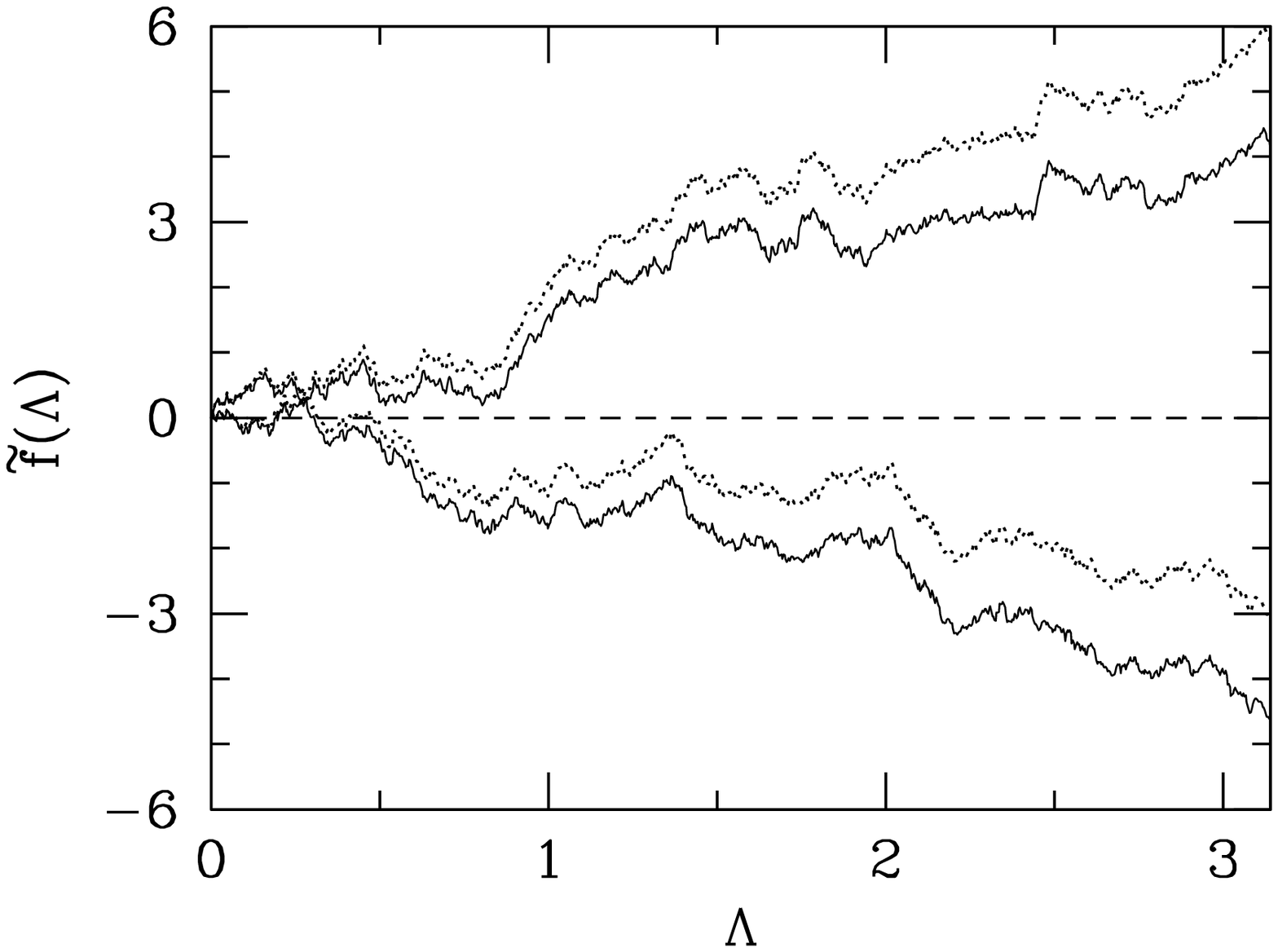,height=9.0cm,width=9.0cm}}
\vspace{-9.0cm}
\centerline{\hspace{+9.0cm}
\psfig{figure=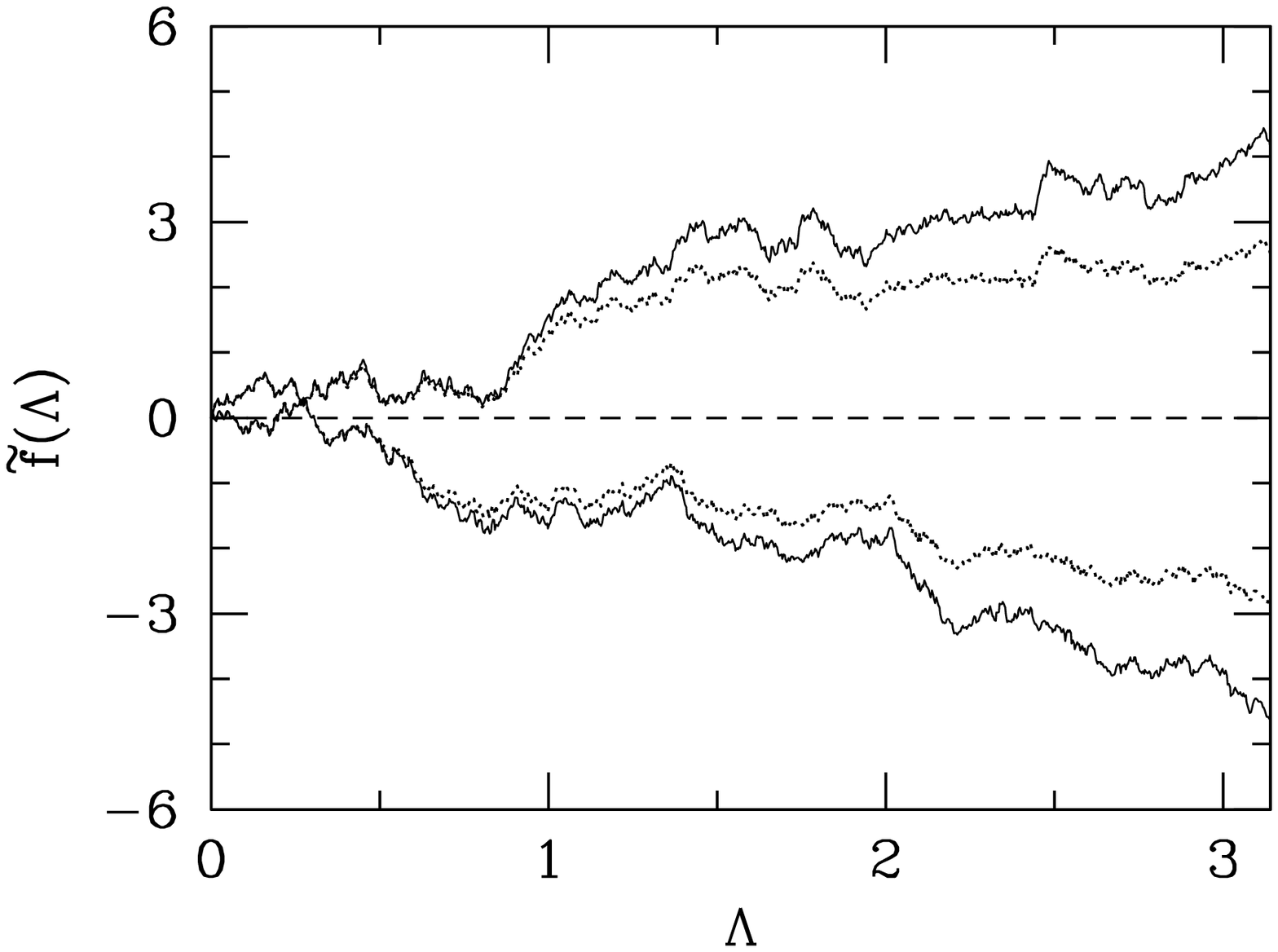,height=9.0cm,width=9.0cm}}
\vspace{-1.4cm}
\caption{\small Sketch of the large-scale behaviour of sample paths of
low-order stochastic diffusion processes. Continuous lines: sample
paths of a Wiener process with no drift term and the diffusion
parameter $A_2=1$. Dotted lines: sample paths of diffusion processes
with the same diffusion parameter but nonzero drift terms; left panel:
order-0 process with $A_1=-0.5$ (Eq. \ref{PATHS}). Right panel:
order-1 process with $A_1=0.2\tilde{f}$.}
\label{FIG_PATHS}
\end{figure}

The It\^o stochastic differential equation cannot be solved in
general. However, as a guideline consider equilibrium solutions with
$\partial\Pi/\partial\Lambda=0$. Simple solutions of (\ref{FP2}) exist
if $A_2$ is independent of $\tilde{f}$ and $\Lambda$, that is,
$A_2=D={\rm const}$. For this choice the correct form of $A_1$ is
needed which is consistent with steady state. For $A_1$ independent of
$\Lambda$, $A_1<0$ would lead to $\tilde{f}$ values `accelerated' to
large $|\tilde{f}|$ with increasing $\Lambda$. Additional dependencies
of $A_1$ on {\it even} powers of $\tilde{f}$ would not change the
situation. However, if $A_1$ depends on {\it odd} powers of
$\tilde{f}$, $A_1>0$ would lead for positive $\tilde{f}$ to be less
positive and for negative $\tilde{f}$ to be less negative. Expanding
$A_1$ in odd powers of $\tilde{f}$ thus yields stable deterministic
states. In this respect important cases are the order-$\nu$ processes,
\begin{equation}\label{ORDERNU}
d\tilde{f}(\Lambda)\,=\,-A\tilde{f}^\nu
d\Lambda\,+\,\sqrt{D}\,dw(\Lambda)\,,\quad\quad
(\nu=0\,\,{\rm or}\,\,{\rm odd})\,.
\end{equation}
For $\nu\ge 3$, hitherto only small-noise expansions are known (e.g.,
Gardiner 1997, p.\,185). We are thus left with the {\it Ansatz}
$\nu=0$ and $\nu=1$. Some historical notes on the resulting equations
can be found in Wax (1954). Although it is not possible to give
graphical presentations of sample paths of stochastic diffusion
processes, Fig.\,\ref{FIG_PATHS} tries to compare trajectories of
order-0 and order-1 diffusion processes obtained with
\begin{equation}\label{PATHS}
\tilde{f}(\Lambda_i)\,=\,-A_1[\tilde{f}(\Lambda_{i-1})]\,\Lambda_i\,+\,
\sqrt{A_2[\tilde{f}(\Lambda_{i-1})]}\,\,w_i\quad (i=1,\ldots,N)\,,
\end{equation}
the $i$\,th resolution variable, $\Lambda_i$, and $w_i$ from Wiener's
construction as described in It\^o \& McKean (1974),
\begin{equation} \label{BROWN}
w_i\,=\,\frac{\Lambda_i}{\sqrt{\pi}}\,g_0\,+\,\sum_{n=1}^{m}\,\sum_{k=2^{n-1}}^{2^n-1}\,
\sqrt{\frac{2}{\pi}}\,\frac{\sin(k\Lambda_i)}{k}\,g_k\quad\quad(0\le\Lambda_i\le\pi)\,,
\end{equation}
where $g_k$, $k\ge 0$ are independent and Gaussian distributed with
zero mean and variance $1$.  For $m\rightarrow\infty$,
Eq. (\ref{BROWN}) converges uniformly in $0\le \Lambda_i\le \pi$ with
probability 1 and is a standard Brownian motion. The sample paths in
Fig.\,\ref{FIG_PATHS} are computed with $m=10$ and $N=1000$ in the
convergence range (larger $m$ values do not significantly improve the
figure). For the order-0 process we set
$A_1[\tilde{f}(\Lambda_{i-1})]\,=-\mu$, for the order-1 process
$A_1[\tilde{f}(\Lambda_{i-1})]=A\tilde{f}(\Lambda_{i-1})$, and
$A_2[\tilde{f}(\Lambda_{i-1})]=1$ for both (see Fig.\,\ref{FIG_PATHS}
for more details).  Fig.\,\ref{FIG_PATHS} illustrates the exceedingly
irregular motion of order-0 processes with an almost infinite `speed',
when $\tilde{f}$ and $\Lambda$ are regarded as the position and time
of a Brownian particle, respectively. Order-0 processes have no
stationary distributions, i.e., as $\Lambda\rightarrow\infty$, any
finite point moves to infinity with probability 1. Order-1 processes
include a linear drift term, which acts like dynamical friction, and
which yields the only Gaussian Markov process in one real variable
with a stationary distribution. The process is often used to model a
more {\it realistic} Brownian motion, in which $\tilde{\Lambda}$ and
$\tilde{\Lambda}'$ are significantly correlated if
$|\Lambda-\Lambda'|\sim 1/A$ (Eq. \ref{OUA4} in Appendix A).

\begin{figure}
\vspace{0.0cm}
\centerline{\hspace{-1.0cm}
\psfig{figure=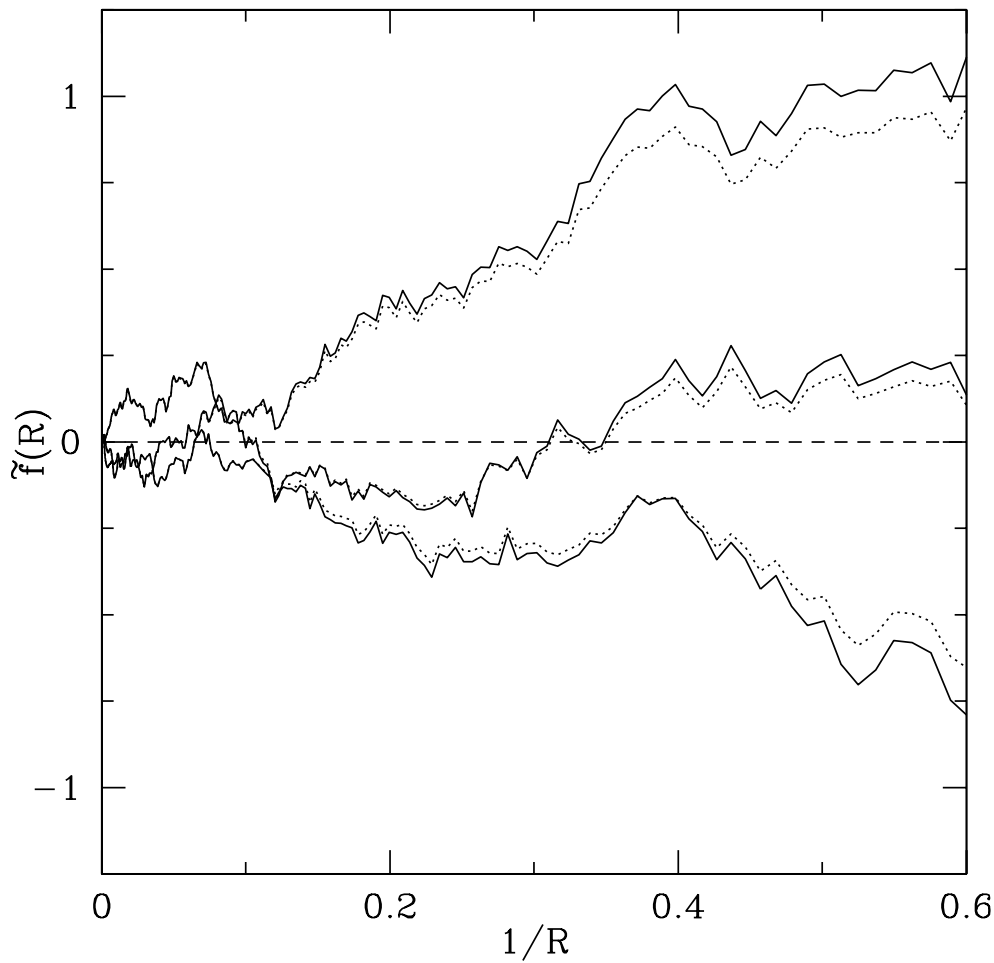,height=13.0cm,width=13.0cm}}
\vspace{0.0cm}
\caption{\small Comparison of sample paths obtained for a high-density
region (upper two trajectories), for a low-density region (lower two
trajectories), and for an average-density region (middle two
trajectories) by filtering a one-dimensional Gaussian random field
(spectral index $n=0$) using a sharp $k$-space filter (continuous
lines) and a truncated exponential filter (dotted lines) with
$A=0.5$. The filter scale, $R$, is set to arbitrary physical
units. The systematic differences between the trajectories obtained
with the sharp $k$-space filter and with the truncated exponential
filter reflect the differences between the trajectories of order-0 and
order-1 diffusion processes shown in Fig.\,\ref{FIG_PATHS}.}
\label{FIG_GAUSS_PATHS}
\end{figure}

In Sect.\,\ref{DPMF} it will be shown that each filter used to smooth
Gaussian matter distributions and to extract masses of isolated
objects is closely related to a specific diffusion process. To
illustrate this point, sample paths are computed for specific
locations (a high, average, low-density region) within a
one-dimensional Gaussian random field which is smoothed with filter
functions where the corresponding diffusion processes are
known. Fig.\,\ref{FIG_GAUSS_PATHS} shows this for the sharp $k$-space
filter giving sample paths which correspond to order-0 diffusion
processes (see Sect.\,\ref{MASS_WIEN}), and for the truncated
exponential filter giving sample paths which correspond to order-1
diffusion processes (see Sect.\,\ref{MASS_OU}). Despite the artificial
smoothing of trajectories at large $1/R$ caused by the limited
numerical resolution in this scale range, the sample paths of the
mathematical models of order-0 and order-1 diffusion processes shown
in Fig.\,\ref{FIG_PATHS} and the sample paths obtained from filtered
Gaussian random fields obtained with the process-specific filter
functions shown in Fig.\,\ref{FIG_GAUSS_PATHS} look qualitatively
quite similar. It is thus seen that order-0 and order-1 diffusion
processes are supposed to give a realistic description of the complex
behaviour of the density contrasts at a given location in a Gaussian
random field when filtered with process-specific filter functions and
different filter widths.

Concerning the physical meaning of the diffusion and drift parameter
it is important to realize that whereas the diffusion parameter
basically determines the relation between immediately adjacing mass
scales, the drift parameter determines the relation between largely
differing mass scales.

Ignoring for the moment the deterministic drift subprocesses, the
drift-less diffusion process is assumed to be independent of mass and
density contrast. The resulting random behaviour of the filtered
density contrasts is thus determined by the inclusion or exclusion of
a new independent $k$ shell. Because of the specific choice of the
resolution variable (see Eq. \ref{RESOLVAR}) and the normalization of
the filter (see Sects.\,\ref{MASS_WIEN_2} and \ref{MASS_OU_2}), the
variance of the Gaussian noise process and thus of the value of the
diffusion parameter, $A_2$, is per construction always unity. With the
addition of a deterministic drift subprocess the idealistic pure
diffusion process can be converted into a more realistic process
because the idealized assumptions of the Wiener process can be
relaxed.

The physical meaning of the drift parameter becomes clear from the
deterministic motions induced (see Figs.\,\ref{FIG_PATHS} and
\ref{FIG_GAUSS_PATHS}). As an example, for $\Lambda$-independent drift
parameters (the diffusion processes discussed here have this property)
the potential
\begin{equation}\label{POT}
U(\tilde{f})\,=\,\int^{\tilde{f}}A_1(\tilde{f}')\,d\tilde{f}'
\end{equation}
might be defined (e.g., Risken 1984, Chap.\,5), leading for the
order-1 process with $A>0$ to a parabolic potential which restricts
the otherwise freely streaming trajectories to smaller values of
$|\tilde{f}|$ (see right panel in Fig.\,\ref{FIG_PATHS}). In
Appendix\,A (Eq. \ref{OUA4}) it is shown that the corresponding
`external deterministic force' introduces realistic nonzero
covariances between different $\tilde{f}(\Lambda)$, that is, it
restricts the trajectories to a smaller $\tilde{f}$ interval around
$\tilde{f}=0$ which has the following two related consequences: (1) It
results in a weakening of the sharp $k$-space filter at larger
comoving wavenumbers $k$ (see Sect.\,\ref{MASS_OU_2}). This is of
great cosmological significance because the corresponding filters in
configuration space are comparatively smooth (see
Fig.\,\ref{FIG_FILT}) and thus frame a more realistic region of
material that immersed into a particular halo. (2) It results in a
correlation between different mass scales and thus effectively weakens
the strict Markovian property, however, without leaving the general
framework of Markovian processes. These correlations are expected to
give more realistic formation histories and biasing schemes compared
to the original Press-Schechter approach.

On the other side, order-0 processes with lead to systematic shifts of
the trajectories in the direction of smaller $\tilde{f}$ or larger
values (left panel in Fig.\,\ref{FIG_PATHS}). This might be
interpreted as a $\Lambda$ (mass)-dependent change of the critical
density threshold, $f_{\rm c}$, which will be discussed in more detail
in Sect.\,\ref{MASS_WIEN_1}.

\section{Diffusion processes and mass functions}\label{DPMF}

In the following the general relation between diffusion processes and
mass functions is developed. As already stated in Sect.\,\ref{EXCSET}
the excursion set formalism assumes that the cumulative mass fraction
$n(>M)/n_{\rm tot}$, where $n_{\rm tot}$ is some normalization
constant, is identified with the fraction of volume elements above
$f_{\rm c}$ when smoothed with a filter of radius $\ge R(M)$ which is
directly related to $\Lambda(R)$ (see below). At least for the high
masses it seems reasonable to associate the change in total volume of
the excursion region with the net mass in objects of this scale (see
Bond et al. 1991, and the useful graphical visualisations given
therein).

In the language of diffusion processes described in
Sect.\,\ref{GENERAL}, the differential mass function is proportional
to the probability that a particular trajectory is absorbed at the
threshold $f_{\rm c}$ in a given $\Lambda$, i.e., mass range. It is
thus necessary to investigate the first passage $\Lambda$ distribution
of the trajectories supposing an absorbing barrier at $f_{\rm
c}>\tilde{f}_0$. Let $\Pi(\tilde{f}_0, \tilde{f},\Lambda)$ be the
probability density to have trajectories at $\tilde{f}(\Lambda)$ {\it
and} that no trajectory has reached $f_{\rm c}$ within the interval
$(0,\Lambda)$. Notice the difference in the definition of $\Pi$ here
and in Sect.\,\ref{GENERAL}. Hence, the probability that absorption
has {\it not} yet occured at $\Lambda$ is (sometimes the $\tilde{f}_0$
variable will be omitted in the argument lists of $F$, $\Pi$, and $g$)
\begin{equation}\label{DPMF1}
F(\tilde{f}_0,f_{\rm c},\Lambda)\,=\,\int^{f_{\rm
c}}_{-\infty}\,\Pi(\tilde{f}_0,\tilde{f},\Lambda)\,d\tilde{f}\,.
\end{equation}
The lower limit, $\tilde{f}=-1$, must be extended formally to
$-\infty$ in order to fulfill for all $\Lambda$ or mass scales under
consideration the assumption of a Gaussian diffusion process. Notice
that $F(\tilde{f}_0,\tilde{f},\Lambda)$ and
$\Pi(\tilde{f}_0,\tilde{f},\Lambda)$ satisfy
(\ref{FP2},\,\ref{FP2a}). The probability density function of the
first passage of trajectories at $\Lambda$ is determined by the
`loss-rate' of sample paths at $\Lambda$, i.e.,
\begin{equation}\label{DPMF2}
g(\Lambda,f_{\rm
c},\tilde{f}_0)\,=\,-\,\frac{\partial}{\partial\Lambda}
\int^{f_{\rm
c}}_{-\infty}\,\Pi(\tilde{f}_0,\tilde{f},\Lambda)\,d\tilde{f}\,.
\end{equation}
In the following $g(\Lambda,f_{\rm c},\tilde{f}_0)$ is called the
pseudo mass function. The probability density function
$g(\Lambda,f_{\rm c},\tilde{f}_0)$ can be determined by solving
Kolmogorov's forward equation (\ref{FP2}) for
$\Pi(\tilde{f}_0,\tilde{f},\Lambda)$ under appropriate boundary
conditions (mirror image method, see Sect.\,\ref{MASS_WIEN_1} and
\ref{MASS_OU_1}). For $\Lambda$-independent $f_{\rm c}$, a second
method determines the Laplace transform of $g(\Lambda,f_{\rm
c},\tilde{f}_0)$,
\begin{equation}\label{DPMF3}
g^*(\lambda,f_{\rm
c},\tilde{f}_0)\,=\,\int_0^\infty\,e^{-\lambda\Lambda}\,g(\Lambda,f_{\rm
c},\tilde{f}_0)\,d\Lambda\,,
\end{equation}
and solves the Laplace-transformed Kolmogorov's backward equation
(\ref{FP2a}) in conjunction with appropriate boundary conditions
(Laplace transform method, see Sects.\,\ref{MASS_WIEN_1} and
\ref{MASS_OU_1}). The most difficult part of the latter method is the
inversion of (\ref{DPMF3}). To obtain the differential mass function,
\begin{equation}\label{GWIEN91}
n(M)\,=\,\frac{\bar{\rho}_0}{M}\,
\left|\frac{d\Lambda(M)}{dM}\right|\,g(\Lambda,f_{\rm
c},\tilde{f}_0)\,,
\end{equation}
we have to derive $\Lambda(M)$ which is closely related to the filter
function of the diffusion process. For Gaussian matter distributions,
this relation is obtained in the following way. In the first step, the
equivalence between the mass variance (\ref{VARI}) and the variance of
the $\tilde{f}$'s,
\begin{equation}\label{VARI2}
\frac{1}{(2\pi)^3}\,\int_0^\infty\, d^3k\,|W_R(k)|^2\,P(k)\,=\,{\rm
var}[\tilde{f}(\Lambda)]\,,
\end{equation}
is used to derive the Fourier transform of the filter function,
$W_R(k)$.  The variances of the $\tilde{f}$'s can be obtained with the
It\^o formalism in a straightforward way (see, e.g., Appendix A). In
the second step, the resulting window function, $W_R(k)$, is
transformed into configuration space to give $W_R(r)$, from which the
filter volume is determined by $V(R)=1/W_R(r=0)$. The mass associated
with the filter is then obtained from $M(R)=\bar{\rho}_0V(R)$, where
$\bar{\rho}_0$ is the present mean mass density of the
Universe. Notice that this step follows the general conventions as
introduced in Lacey \& Cole (1994). In the last step, the general
relation between $\Lambda$ and $R$ (see Eq. \ref{WIN7}) is
applied to give the wanted $\Lambda(M)$ relation. Sect.\,\ref{SHARP}
presents a discussion of the general $\Lambda(R)$ relation.

\section{The $\Lambda$-$R$ resolution scale relation}\label{SHARP}

Within the framework of stochastic diffusion processes, mass functions
are determined by the probabilities that the continuous sample paths
of the diffusion process are first absorbed at $\tilde{f}=f_{\rm
c}$. Analytical results are obtained by assuming no correlations
between $\tilde{f}$ and its infinitesimal increments, $d\tilde{f}$,
but (see, e.g., Eq. \ref{OUA9} for $A=0$ in Appendix A)
\begin{eqnarray}\label{CORR}
{\rm E}[\tilde{f}(R)\cdot
\tilde{f}(R')]\,=\,\frac{1}{(2\pi)^3}\,\int\,d^3k\,W_R(k)\,W^*_{R'}(k)\,P(k)
\,=\,\sigma^2({\rm max}\{R,R'\})\,.
\end{eqnarray}
The last equality can be achieved if the filter function, $W_R(k)$,
has the form of the sharp $k$-space filter (Eq. \ref{SHARPK}), and
the resolution variable, $\Lambda$, is defined by
$\Lambda=\sigma^2(R)$. This is the prescription of the fundamental
stochastic diffusion process. Hence, the $\Lambda$-dependence of the
probabilities $\Pi(\tilde{f}_0,\tilde{f},\Lambda)$ can be obtained
from the simplified Fokker-Planck equation,
\begin{equation}\label{SFP}
\frac{\partial}{\partial\Lambda}\,
\Pi(\tilde{f}_0,\tilde{f},\Lambda)\,=\,
\frac{D}{2}\,\frac{\partial^2}{\partial\tilde{f}^2}\,
\Pi(\tilde{f}_0,\tilde{f},\Lambda)\,,
\end{equation}
with the diffusion coefficient, $D=1$. It is a classical exercise of
the It\^o stochastic formalism to show that (\ref{SFP}) can be deduced
from (\ref{ITO}) and {\it vice versa} for the constants $A_1=0$ and
$A_2=D$. Although these assumptions (including boundary conditions,
for example, eqs.\,\ref{GWIEN4}) suffice to derive the exact form and
normalization of the Press-Schechter function, they are for several
reasons not very realistic. Zero-correlations between $\tilde{f}$ and
its increments are only attainable if the stochastic force, $K$, has
zero mean and the covariance ${\rm E}[K(\Lambda')\cdot
K(\Lambda)]\,=\,{\rm cov}(\Lambda'-\Lambda)\,=\,D\,\delta_{\rm
D}(\Lambda'-\Lambda)$. This gives a white noise power spectrum and
thus an {\it infinite} variance. As $\Lambda\rightarrow 0$, solutions
of the simplified Fokker-Planck equation tend to the Dirac delta
distribution. As $\Lambda\rightarrow\infty$, the probabilities,
$\Pi(\cdot)$, tend to zero, and no stationary solution exists. It is
seen that in the same sense as the sharp $k$-space filter seems to be
an idealistic choice to solve the mass function problem, the related
fundamental stochastic diffusion process with its vanishing drift term
must also be regarded as idealistic.

No consistent solutions are, however, attainable for $A_1\neq 0$ in
combination with the original choice in the excursion set theory for
the quasi time variable $\Lambda\equiv \sigma^2$, assuming that
$\sigma^2$ is defined via a process-specific filter function and the
power spectrum in the standard way. The inconsistency is seen by
choosing, for example, ${\rm
E}\{\Delta\tilde{f}|\tilde{f}\}=A\,\tilde{f}$ in Eqs. (3.4) of Bond et
al. (1991), where $A$ is a constant independent of $\tilde{f}$ and
$\Lambda$, that is, $A_1\,=\,A\,\tilde{f}$ in (\ref{ITO}). For
$A_2=D$, this is the order-1 or Ornstein-Uhlenbeck stochastic
diffusion process. Appendix\,A summarizes some basic properties of
this diffusion process. Assuming $A>0$, as $\Lambda\rightarrow\infty$
the variances (see Eq. \ref{OUA5}) tend to the finite value,
$\sigma^2=D/(2A)$, but in the excursion set theory it is assumed that
$\sigma^2=\Lambda$ which tends to infinity. This contradiction shows
that the specific choice, $\Lambda=\sigma^2$, is not suitable for a
discussion of more general stochastic diffusion processes.

However, a general resolution scale relation, $\Lambda(R)$, is needed
to handle different diffusion processes in an homogeneous way. The
most natural choice is the $\Lambda(R)$ scale relation defined by the
fundamental process,
\begin{equation}\label{WIN7}
\Lambda(R)\,=\,\frac{1}{2\pi^2}\,\int_0^{\frac{1}{R}}dk\,k^2\,P(k)\,=\,
\frac{1}{2\pi^2}\,\frac{P_0}{n+3}\,R^{-(n+3)}\quad\quad
{\rm for}\,\,\,P(k)=P_0k^n\quad(n>-3)\,,
\end{equation}
as the common resolution scale for {\it all} diffusion processes
investigated. A further motivation for this specific choice is given
by the transformation property of the Wiener process. For example, the
Wiener process, $d\tilde{f}(\Lambda)=dw(\Lambda)$, is converted into
the Ornstein-Uhlenbeck process,
$d\tilde{f}(\Lambda)=-A\tilde{f}d\Lambda+\sqrt{D}dw(\Lambda)$, by the
transformation $\tilde{f}(\Lambda)=h(\Lambda)\tilde{f}_{\rm
w}[T(\Lambda)]$, with the amplitude transformation
$h(\Lambda)=e^{-A\Lambda}$, and the resolution variable transformation
$T(\Lambda)=D/(2A)e^{2A\Lambda}$, where again -- and this is the
important point -- the resolution variable of the fundamental process,
$\Lambda$, sets the basic resolution scale. Notice that the
instrumental distinction between $\Lambda$ and $\sigma^2$ is necessary
because $\Lambda\neq\sigma^2$ in general. Only for the fundamental
(Wiener) process we have $\Lambda=\sigma^2$. In the following sections
pseudo mass functions, $g(\Lambda,f_{\rm c},\tilde{f}_0)$, are derived
and transformed into `true' differential mass functions, $n(M)$, using
the Eqs. (\ref{GWIEN91}, \ref{VARI2}).

\begin{figure}
\vspace{-1.7cm}
\centerline{\hspace{-9.0cm}
\psfig{figure=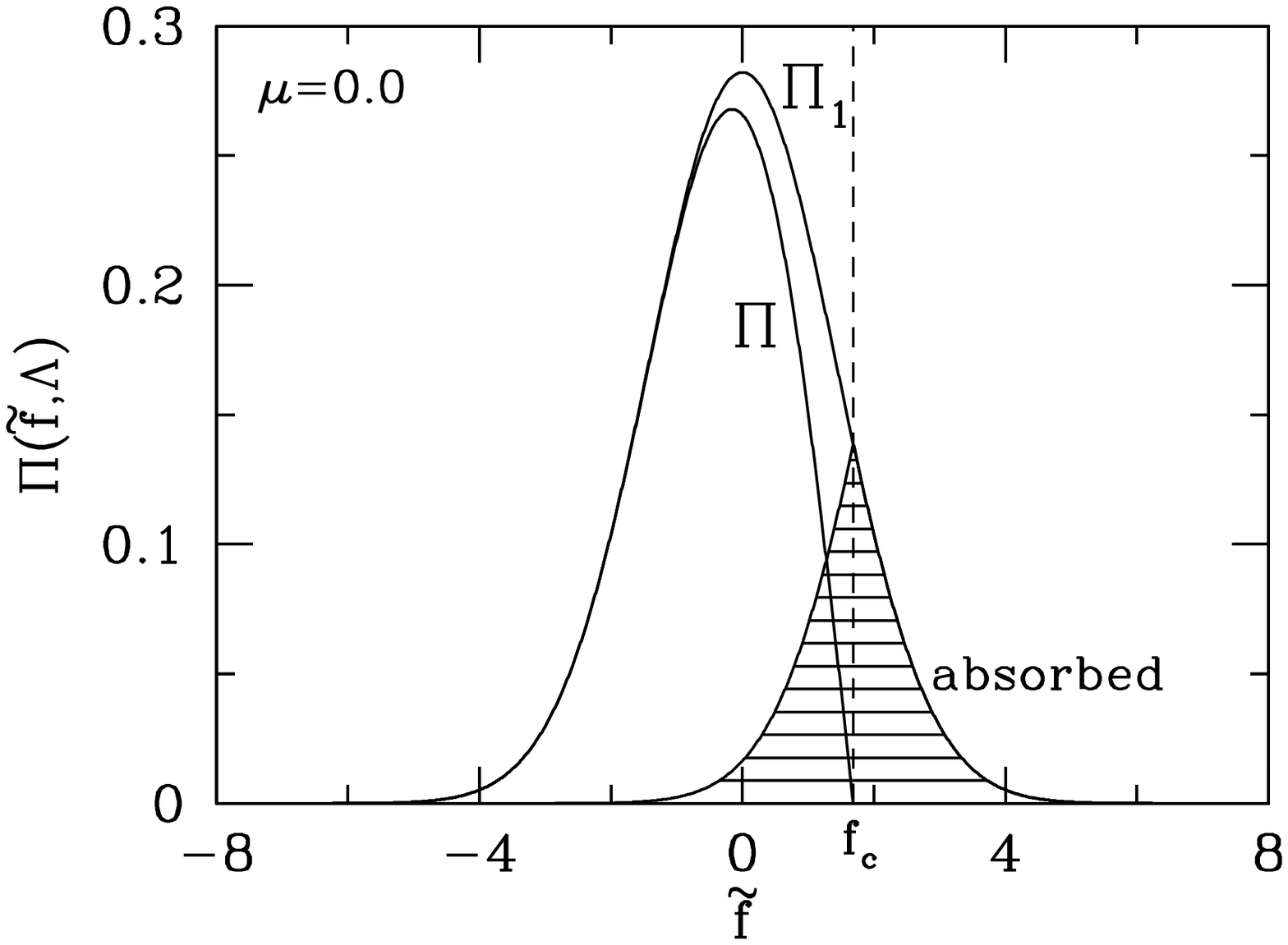,height=9.0cm,width=9.0cm}}
\vspace{-9.0cm}
\centerline{\hspace{+9.0cm}
\psfig{figure=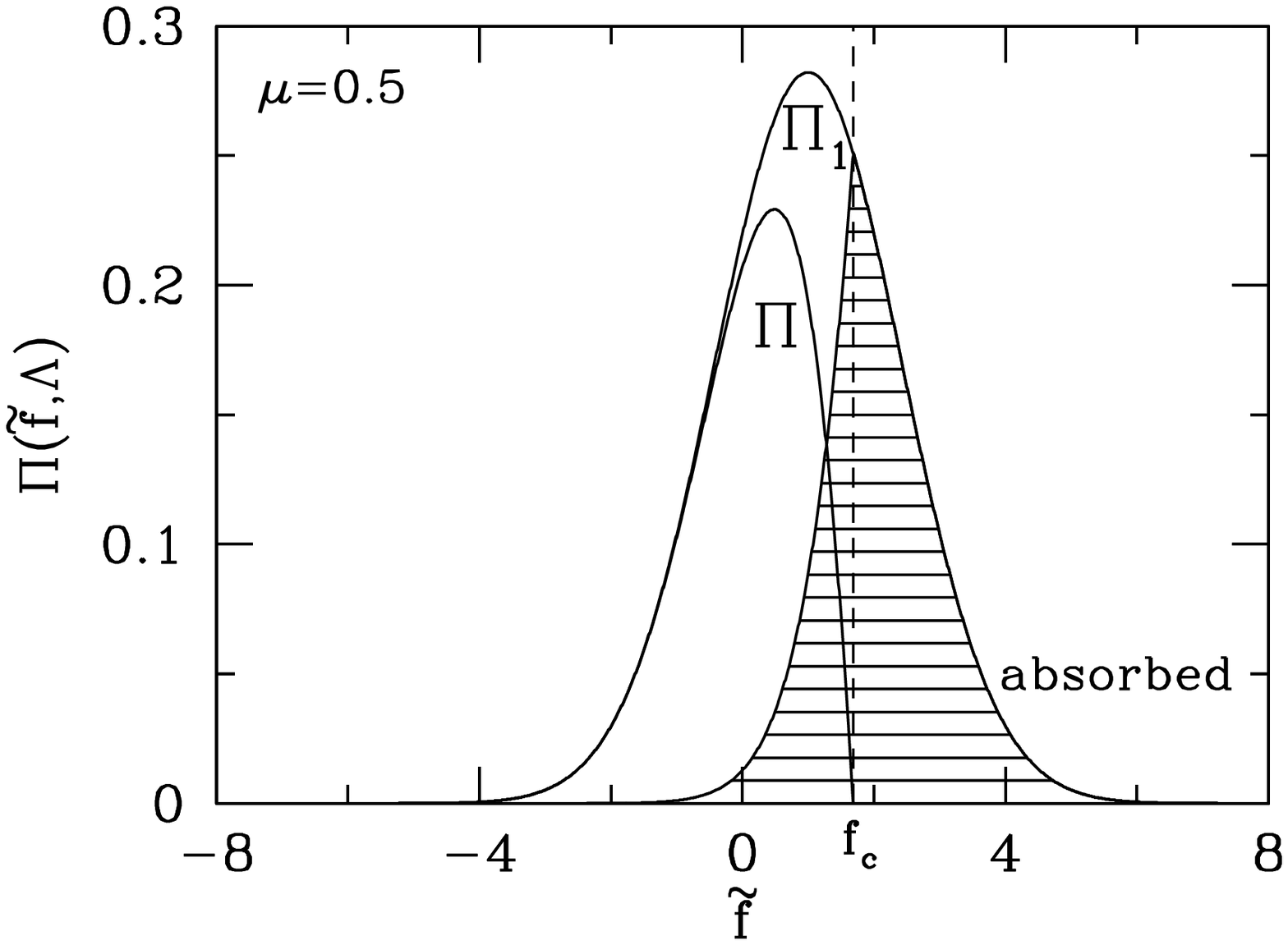,height=9.0cm,width=9.0cm}}
\vspace{-1.5cm}
\caption{\small Probability density distributions for unconstrained
sample paths ($\Pi_1$, see Eq. \ref{GWIEN5}), for nonabsorbed sample
paths ($\Pi$, see Eq. \ref{GWIEN8}), and for absorbed sample paths
(hatched areas) for order-0 processes without drift ($\mu=0$, left
panel) and with the drift parameter $\mu=0.5$ (right panel).}
\label{FIG_PIFL}
\end{figure}

\section{Mass functions from order-0 (Wiener) stochastic diffusion 
processes}\label{MASS_WIEN}

Order-0 Gaussian stochastic diffusion processes are characterized by
the constants $A_1=-\mu$ (or $A=-\mu$ and $\nu=0$ in
Eq. \ref{ORDERNU}), and $A_2=D$ in (\ref{ITO}), used to write the
It\^o stochastic differential equation in the form
\begin{equation}\label{0_ORDER}
d\tilde{f}(\Lambda)\,=\,\mu\,d\Lambda\,+\,\sqrt{D}\,dw(\Lambda)\,,
\end{equation}
with the solution
\begin{equation}\label{GWIEN11}
\tilde{f}(\Lambda)\,=\,\tilde{f}_0\,+\,\mu\Lambda\,+\,
\sqrt{D}\,\int_0^\Lambda\,dw(\Lambda')\,,
\end{equation}
and hence the instantaneous mean ${\rm
E}[\tilde{f}(\Lambda)]\,=\,\tilde{f}_0\,+\,\mu\Lambda$, and the
instantaneous variance ${\rm
var}[\tilde{f}(\Lambda)]\,=\,D\,\Lambda$. The corresponding
Kolmogorov's forward equation is
\begin{equation}\label{GWIEN2}
\frac{D}{2}\,\frac{\partial^2\Pi}{\partial\tilde{f}^2}\,-\,
\mu\,\frac{\partial\Pi}{\partial\tilde{f}}\,=\,
\frac{\partial\Pi}{\partial\Lambda}\quad\quad(-\infty<\tilde{f}< f_{\rm c})\,.
\end{equation}
The filtered density contrasts are restricted to values below the zero
redshift critical overdensity, e.g., $f_{\rm
c}=(3/20)(12\pi)^{(2/3)}\Omega_0^{0.0185}$ for the spherical symmetric
cosmological infall model (Gunn \& Gott 1972, Bertschinger 1985,
Navarro, Frenk \& White 1996). Notice that in (\ref{GWIEN2}), $\Pi$ is
the probability density for $\tilde{f}$ conditional to
$\tilde{f}(0)=\tilde{f}_0$, {\it and} that the process sample paths do
not reach $f_{\rm c}$ in the resolution interval $(0,\Lambda)$. The
cumulative distribution function, $\bar{F}(\tilde{f}_0,f_{\rm
c},\Lambda)=1-F(\tilde{f}_0,f_{\rm c},\Lambda)$, defined by the
probability density of first absorbed sample paths, is directly
related to the pseudo mass function via eqs.\,(\ref{DPMF1}),
(\ref{DPMF2}), and (\ref{GWIEN91}). We thus have
\begin{equation}\label{GWIEN3}
\bar{F}(\tilde{f}_0,f_{\rm
c},\Lambda)\,=\,1\,-\,\int_{-\infty}^{f_{\rm c}}\,
\Pi(\tilde{f}_0,\tilde{f},\Lambda)\,d\tilde{f}\,,
\end{equation}
and compute $\bar{F}(\tilde{f}_0,f_{\rm c},\Lambda)$ solving
(\ref{GWIEN2}) under the boundary conditions
\begin{equation}\label{GWIEN4}
\Pi(\tilde{f},\Lambda=0)\,=\,\delta_{\rm D}(\tilde{f})\,,\quad
\Pi(\tilde{f}=f_{\rm c},\Lambda)\,=\,0\quad\quad(\Lambda>0)\,.
\end{equation}
Whereas the first equation reflects the deterministic initial state of
the diffusion process, the second equation gives the appropriate
boundary condition for an absorbing barrier at $f_{\rm c}$ (e.g., Cox
\& Miller, 1965, p.\,219): Assume a nonvanishing infinitesimal
variance when $\tilde{f}=f_{\rm c}$. The probability that the
trajectory is absorbed within $\Delta\Lambda$ can be approximated by
the pro\-duct of the probability that the trajectory is near (i.e.,
slightly below) $f_{\rm c}$, and the probability that the next
increment $\Delta\tilde{f}$ -- which is proportional to
$\sqrt{\Delta\Lambda}$ (see Sect.\,\ref{GENERAL}) -- carries it beyond
$f_{\rm c}$. However, the latter probability is assumed to be always
larger zero. In order to keep the probability density (i.e., the
absorption probability per $\Delta\Lambda$) finite that the trajectory
is absorbed when $\tilde{f}\rightarrow f_{\rm c}$, the former
probability must be zero because
$\sqrt{\Delta\Lambda}/\Delta\Lambda\rightarrow\infty$ when
$\Delta\Lambda\rightarrow0$.

\subsection{Pseudo mass functions}\label{MASS_WIEN_1}

The unconstrained case of Eq. (\ref{GWIEN2}) has the general
solution (Fig.\,\ref{FIG_PIFL})
\begin{equation}\label{GWIEN5}
\Pi_1(\tilde{f}_0,\tilde{f},\Lambda)\,=\, \frac{1}{\sqrt{2 \pi
D\Lambda}}\,\exp\left[
-\frac{(\tilde{f}-\tilde{f}_0-\mu\Lambda)^2}{2D\Lambda}\right]\,,
\end{equation}
where the sample paths start at $\tilde{f}_0$. Exploiting the symmetry
of the diffusion process, which still holds for $\mu\neq 0$, a
solution is obtained by placing one source of sample paths at
$\tilde{f}_0=0$, and a second source at $\tilde{f}_0=2f_{\rm c}$
(mirror image method), that is,
$\Pi(\tilde{f},\Lambda)\,=\,\Pi_1(0,\tilde{f},\Lambda)\,+\,
\alpha\,\Pi_1(2f_{\rm c},\tilde{f},\Lambda)$, where the factor
$\alpha$ is determined by the condition $\Pi(f_{\rm
c},\Lambda)=0$. The solution gives the probability of nonabsorbed
sample paths and takes the form
\begin{eqnarray}\label{GWIEN8}
\Pi(\tilde{f},\Lambda)\,=\, \frac{1}{\sqrt{2\pi D\,\Lambda}}\,\left\{
\exp\left[-\frac{(\tilde{f}-\mu\Lambda)^2}{2D\Lambda}\right]
\,-\,\exp\left(\frac{2\mu f_{\rm c}}{D}\right)\,\,
\exp\left[-\frac{(\tilde{f}-2f_{\rm
c}-\mu\Lambda)^2}{2D\Lambda}\right]\right\}\quad\quad(-\infty<\tilde{f}<f_{\rm
c})\,.
\end{eqnarray}
It is seen that the second source in (\ref{GWIEN8}), i.e., the last
term on the right-hand side, has a negative amplitude and acts as a
`sink' for sample paths. The drift parameter, $\mu$, introduces a mass
($\Lambda$)-dependent shift of $\Pi_1$ which increases ($\mu>0$) or
decreases ($\mu<0$) the probability for absorbed sample paths (see
shaded regions in Fig.\,\ref{FIG_PIFL}). Although the threshold,
$f_{\rm c}$, is assumed to be mass-independent, the systematic drift
might be interpreted as an effective change of the critical threshold
in a mass-dependent way. Notice that a possible physical
interpretation is that for high mass (low-$\Lambda$) clumps the effect
of the surrounding volume elements is small so that the collaps will
be spherical along all three directions with $f_{\rm c}$ close to the
spherical collaps model (Bernadeau 1994). For small mass (high
$\Lambda$) clumps the effects of the surrounding volume elements might
be more significant and can thus alter the effective value of $f_{\rm
c}$.

\begin{figure}
\vspace{-2.7cm}
\centerline{\hspace{-9.0cm}
\psfig{figure=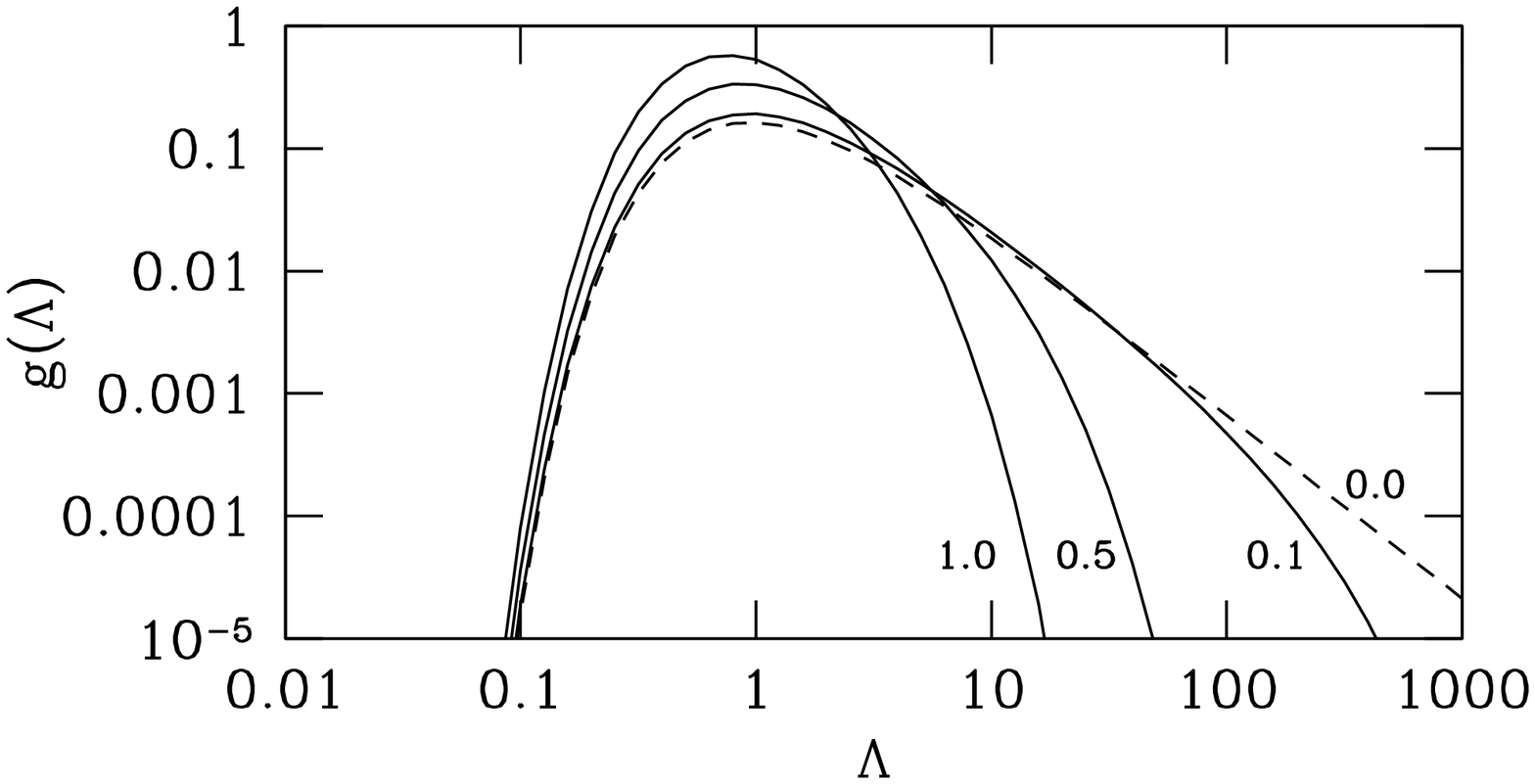,height=9.0cm,width=9.0cm}}
\vspace{-9.0cm}
\centerline{\hspace{+9.0cm}
\psfig{figure=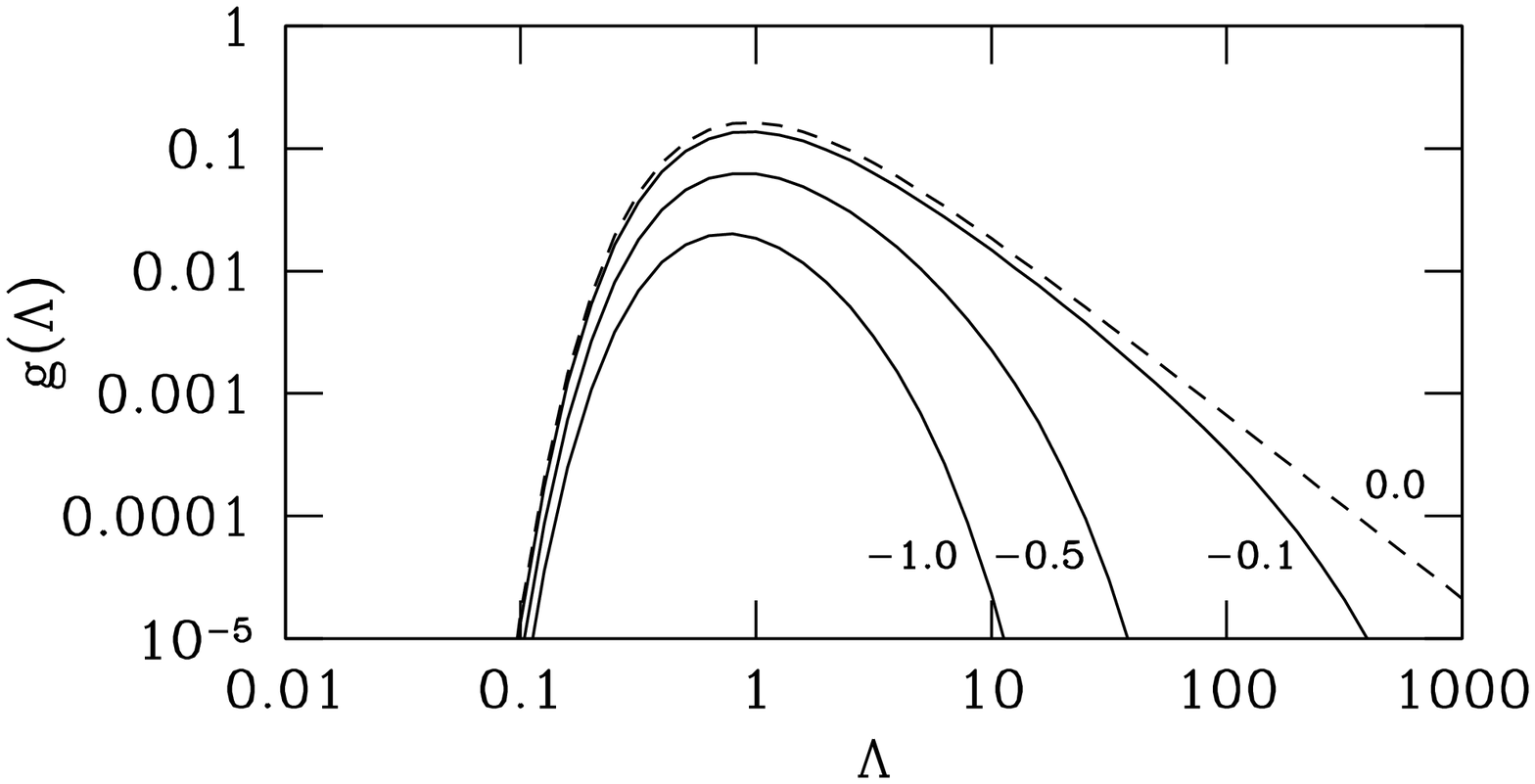,height=9.0cm,width=9.0cm}}
\vspace{-1.8cm}
\caption{\small Pseudo mass functions from order-0 processes for
different $\mu$ values (continuous lines). For comparison, the
Press-Schechter pseudo mass function ($\mu=0$) is given (dashed
lines). Both the mirror image and the Laplace transform method give
the same results.}
\label{FIG_W}
\end{figure}

Substituting (\ref{GWIEN8}) into (\ref{DPMF2}) gives the pseudo mass
function for the order-0 diffusion process,
\begin{equation}\label{GWIEN9}
g(\Lambda,f_{\rm c})=\,\frac{\bar{F}(f_{\rm
c},\Lambda)}{d\Lambda}\,=\, \frac{f_{\rm c}}{\sqrt{2\pi D}}\,
\left(\frac{1}{\Lambda}\right)^{\frac{3}{2}}\,
\exp\left[-\frac{(f_{\rm c}-\mu\Lambda)^2}{2D\Lambda}\right]\,.
\end{equation}
The mirror image method is widely used in solving problems of heat
conduction and heat diffusion (e.g., Landau \& Lifschitz 1966). It
`reduces' the solution of the diffusion problem originally restricted
to a specific domain of diffusion space to an infinite region.

Since the absorbing barrier is $\Lambda$-independent, exact pseudo
mass functions can also be deduced by taking Laplace transforms in the
Kolmogorov's backward equation (Laplace transform method) giving the
partial differential equation for $g^*(\lambda,f_{\rm c},\tilde{f}_0)$,
\begin{equation}\label{GWIEN9b}
\frac{D}{2}\,\frac{\partial^2}{\partial\tilde{f}_0^2}\,
g^*(\lambda,f_{\rm
c},\tilde{f}_0)\,+\,\mu\,\frac{\partial}{\partial\tilde{f}_0}\,
g^*(\lambda,f_{\rm c},\tilde{f}_0)\,=\,\lambda\,g^*(\lambda,f_{\rm
c},\tilde{f}_0)\,.
\end{equation}
The backward equation is the appropriate equation because of the fixed
final state $f_{\rm c}$, we seek for solutions $g^*$ as a function of
the backward variable $\tilde{f}_0$ and set $\tilde{f}_0=0$ later. For
order-0 diffusion processes the general solution of (\ref{GWIEN9b}) is
\begin{equation}\label{GWIEN9a}
g^*(\lambda,f_{\rm
c},\tilde{f}_0)\,=\,Ae^{\tilde{f}_0H_1(s)}\,+\,Be^{\tilde{f}_0H_2(s)}\,.
\end{equation}
Inserting (\ref{GWIEN9a}) in (\ref{GWIEN9b}) gives a quadratic
equation for the $H$'s, and thus, after applying the condition
$g^*(\lambda,f_{\rm c},\tilde{f}_0)\le 1$ and $g^*(\lambda,f_{\rm
c},\tilde{f}_0=f_{\rm c})=1$ (immediate absorption) from which the
constants $A$ and $B$ are determined, the Laplace transform of the
pseudo mass function (for $\tilde{f}_0=0$),
\begin{equation}\label{LPT1}
g^*(\lambda,f_{\rm c})\,=\, \exp\left[\frac{f_{\rm
c}}{D}\left(\mu-\sqrt{\mu^2+2D\lambda}\right)\right]\,.
\end{equation}
Inversion of (\ref{LPT1}) again gives (\ref{GWIEN9}): both the mirror
image and the Laplace transform method lead to the same result.  For
$\mu\rightarrow 0$, Eq. (\ref{GWIEN9}) is the well-known
Press-Schechter pseudo mass function with the correct normalization
(see Fig.\,\ref{FIG_W}).

\begin{figure}
\vspace{
0.0cm}\hspace{3.0cm}\psfig{figure=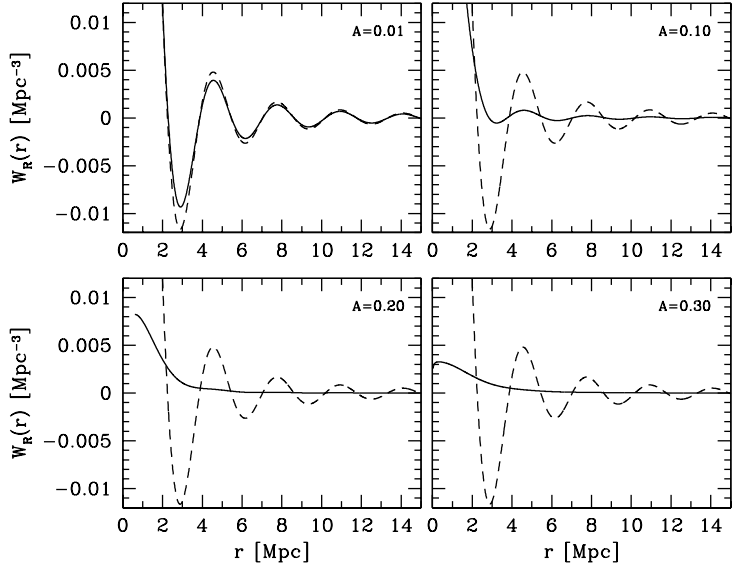,height=10cm,width=10cm}
\vspace{-2.2cm}
\caption{\small Filter functions in configuration space: the sharp
$k$-space filter (dashed lines), and the truncated exponential filter
(continuous lines) for the filter scale, $R=0.25\,{\rm Mpc}$. The
drift parameters have values between zero (dashed line) and 0.3. The
amplitude of the power spectrum is $P_0=200\,{\rm Mpc}$, and the
spectral index $n=-2$ (normalization and shape similar to a CDM power
spectrum). The normalized Hubble constant, $h=0.5$, is used in units
of $H_0=100\,{\rm km}\,{\rm s}^{-1}\,{\rm Mpc}^{-1}$.}\label{FIG_FILT}
\end{figure}

\subsection{Mass functions}\label{MASS_WIEN_2}

\begin{figure}
\vspace{-1.3cm}
\centerline{\hspace{-9.5cm}
\psfig{figure=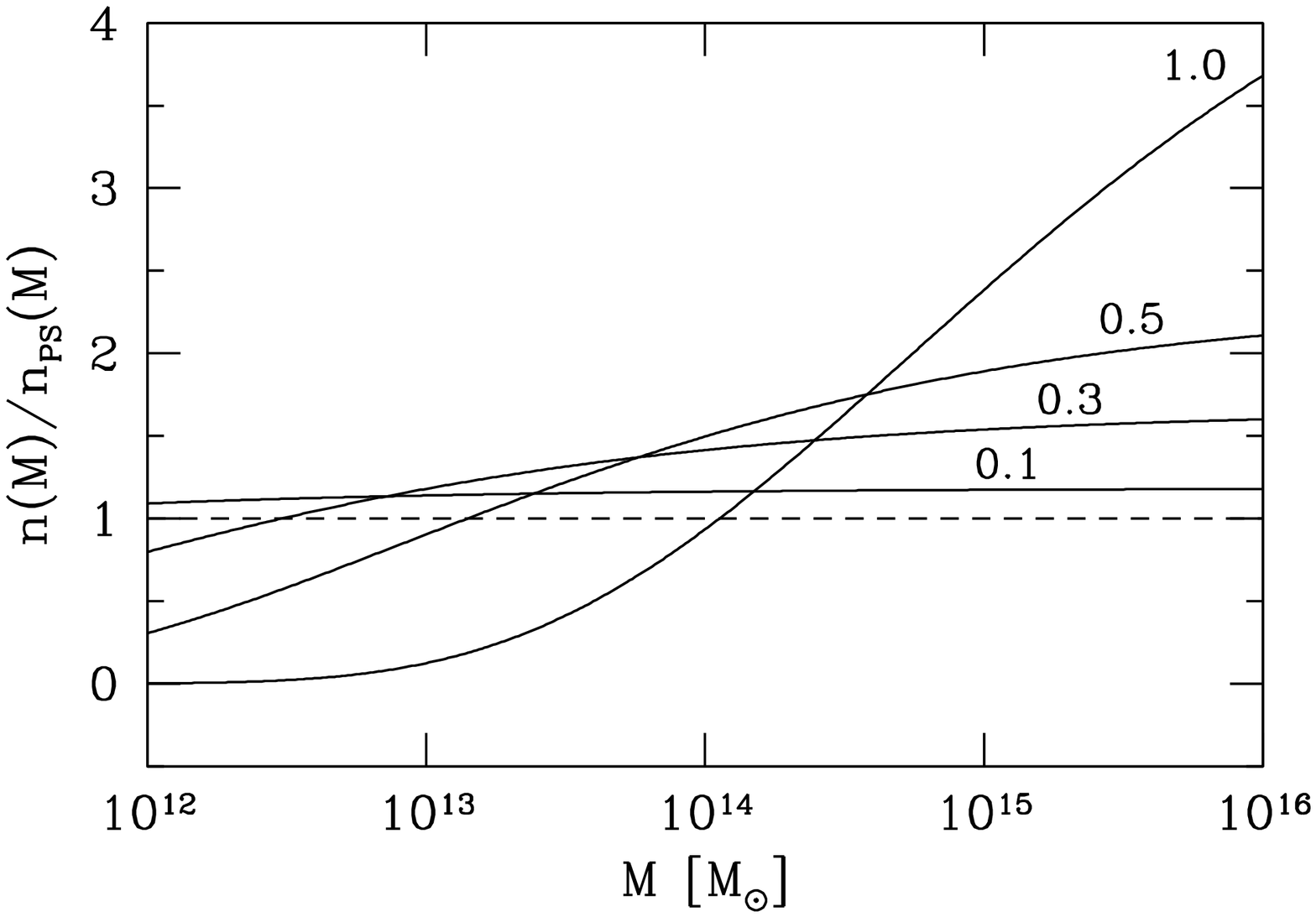,height=9.0cm,width=9.0cm}}
\vspace{-9.0cm}
\centerline{\hspace{+9.0cm}
\psfig{figure=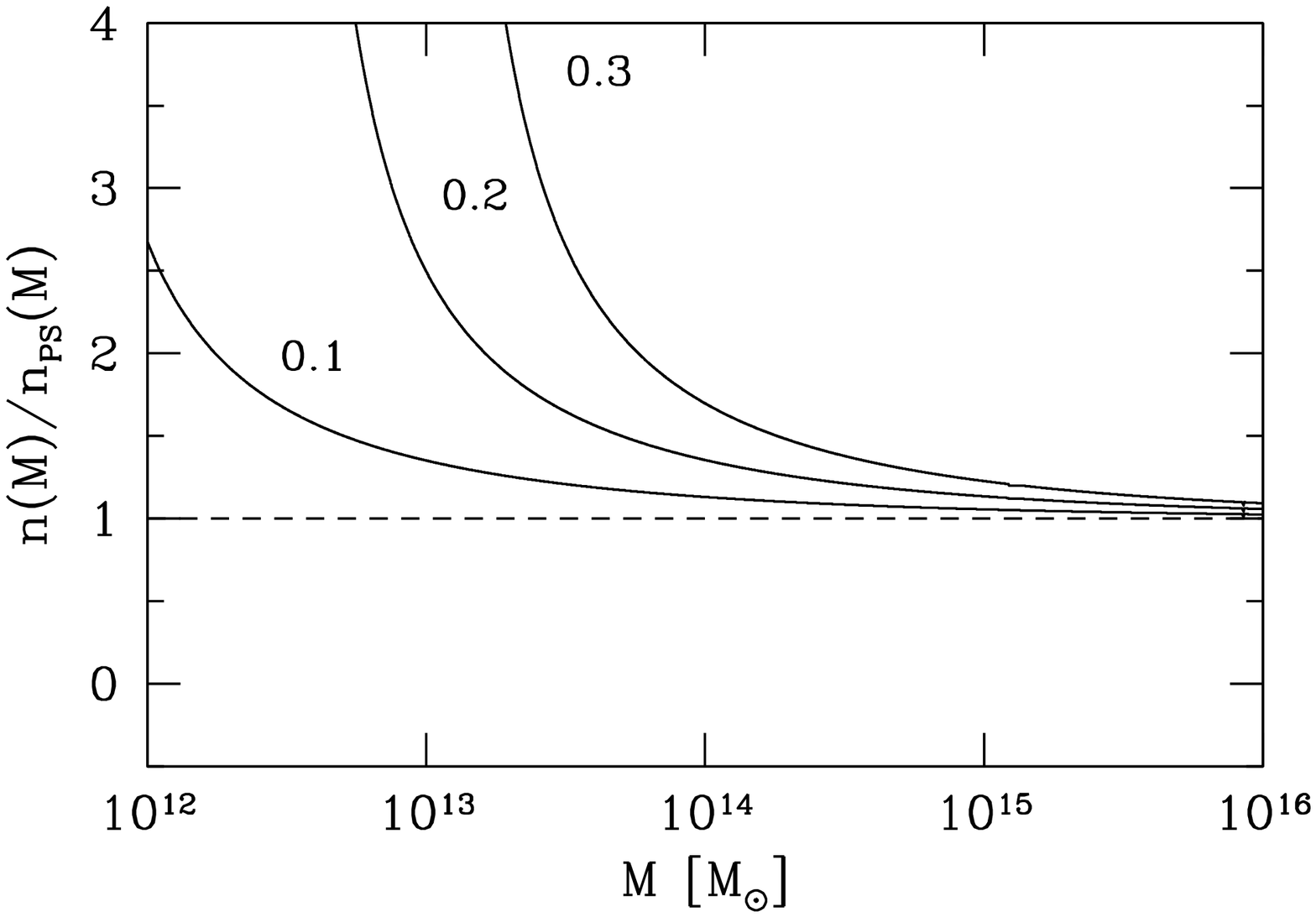,height=9.0cm,width=9.0cm}}
\vspace{-1.5cm}
\caption{\small Ratios of the mass functions induced by order-0 (left)
and order-1 processes (right). A scale-invariant power spectrum with
the index $n=-2$ and the amplitude $P_0=200\,{\rm Mpc}$, the critical
density contrast, $f_{\rm c}=1.68$, the normalized Hubble
constant, $h=0.5$, and the density parameter, $\Omega_0=1.0$, are
used. The different curves are labeled by the value of the
corresponding drift parameter. Masses are given in solar
units.}\label{FIG_NM_W_OU}
\end{figure}

Equating the process variance, $D\Lambda$, of the $\tilde{f}$'s with
the mass variance, $\sigma^2(R)$, and using (\ref{WIN7}) gives the
relation which determines the filter function for this process,
\begin{equation}\label{WIN11}
\frac{1}{2\pi^2}\,\int_0^\infty dk\,k^2\,|W_R(k)|^2\,P(k)\,=\,
D\,\Lambda \,=\,\frac{D}{2\pi^2}\,\int_0^{\frac{1}{R}}dk\,k^2\,P(k)\,.
\end{equation}
As expected, the filter function is the sharp $k$-space filter
(Eq. \ref{SHARPK}). The normalization of the filter function yields
$D=1$. Note that the relation between $\Lambda$ and $\sigma^2$ is
simply given by $\sigma^2(\Lambda)\,=\,\Lambda$. In configuration
space, the volume and the mass within the filter radius $R$ as well as
the sharp $k$-space filter function (Fig.\,\ref{FIG_FILT}) are
\begin{equation}\label{WIN71a}
V_{\rm S}\,=\,\frac{1}{W_{R}(r=0)}\,=\,6\pi^2\,R^3\,,\quad\quad
M_{\rm S}\,=\,6\pi^2\,\bar{\rho}_0\,R^3\,,
\end{equation}
\begin{equation}\label{WIN71}
W_{R}(r)\,=\,\frac{1}{2\pi^2r^3}\,\left[\sin\left(\frac{r}{R}\right)\,-\,
\left(\frac{r}{R}\right)\cos\left(\frac{r}{R}\right)\right]\,.
\end{equation}
For a scale-invariant power spectrum, $P(k)=P_0k^n$, the $M(\Lambda)$
relation becomes
\begin{equation}\label{WIN73}
M_{\rm S}\,=\,6\pi^2\,\bar{\rho}_0\,
\left[\frac{P_0}{2\pi^2\,(n+3)\,\Lambda}\right]^{\frac{3}{n+3}}\,,
\end{equation}
The condition for hierarchical growth of structure (variance decreases
with mass) is $n>-3$. Notice that $\Lambda$ is basically determined by
the filter scale $R$, and by the power spectrum of the mass density
distribution. This has important implications when mass functions,
$n(M)$, are derived by using Eq. (\ref{GWIEN91}). The $\Lambda(M)$
relation (we omit the subscript of $M$),
\begin{equation}\label{WIN742a}
\Lambda(M)\,=\,\frac{P_0}{2\pi^2(n+3)}\,\,
\left(\frac{6\pi^2\bar{\rho}_0}{M}\right)^{\frac{n+3}{3}}\,,
\end{equation}
where $\bar{\rho}_0=2.7755\times 10^{11}\,\Omega_0 h^2\,{\rm
M}_{\odot}\,{\rm Mpc}^{-3}$, gives the mass function,
\begin{equation}\label{WIN742}
n(M)\,=\,\frac{(2\pi^2)^{\frac{3}{n+3}}}{18\pi^2\sqrt{2\pi}}\,
(n+3)^{\frac{6+n}{3+n}}\,\frac{f_{\rm
c}}{M\,P_0^{\frac{3}{n+3}}}\,
\Lambda^{\frac{3-n}{6+2n}}\,\exp\left[-\frac{(f_{\rm
c}-\mu\Lambda)^2}{2\Lambda}\right]\,.
\end{equation}
Ratios of the mass functions induced by order-0 processes with
positive drift parameters to the Press-Schechter mass function are
shown in Fig.\,\ref{FIG_NM_W_OU}. Eq. (\ref{WIN742}) differs from the
standard Press-Schechter function by the extra $\mu\Lambda$ term in
the exponential. As expected positive drift parameters increase the
fraction of high $\tilde{f}$ and hence the number of high-mass objects
to the expense of objects with lower mass. For $\mu\rightarrow 0$,
Eq. (\ref{WIN742}) converges to the Press-Schechter mass function.

\section{Mass functions from order-1 (Ornstein-Uhlenbeck) stochastic 
diffusion processes}\label{MASS_OU}

Order-1 Gaussian stochastic diffusion processes are characterized by
the parameters $A_1=A\tilde{f}$ (or $\nu=1$ in Eq. \ref{ORDERNU}),
and $A_2=D$, used to write the It\^o stochastic differential equation
in the form
\begin{equation}\label{OU2}
d\tilde{f}(\Lambda)\,=\,-A\,\tilde{f}\,d\Lambda\,+\,\sqrt{D}\,dw(\Lambda)\,.
\end{equation}
Historical notes on this equation can be found in Wax (1954). Some
basic informations about this process are given in Appendix A. The
formal solution for $\tilde{f}_0=0$ is
\begin{equation}\label{OUA3a}
\tilde{f}(\Lambda)\,=\,\sqrt{D}\,\int_0^{\Lambda}e^{-A(\Lambda-\Lambda')}\,dw(\Lambda')\,,
\end{equation}
yielding the instantaneous mean, ${\rm E}[\tilde{f}(\Lambda)]=0$, and
the instantaneous variance,
\begin{equation}\label{OU2a}
{\rm
var}[\tilde{f}(\lambda)]\,=\frac{D}{2A}\,\left(\,1\,-\,e^{-2A\Lambda}\,\right)\,.
\end{equation}
It is straightforward to write down the Kolmogorov's forward and
backward equations (\ref{FP2},\,\ref{FP2a}) for this process. The
boundary conditions are identical to those already used for the
order-0 processes in Sect.\,\ref{MASS_WIEN}

\subsection{Pseudo mass functions}\label{MASS_OU_1}

\begin{figure}
\vspace{-3.7cm}
\centerline{\hspace{-9.0cm}\hspace{9.0cm}
\psfig{figure=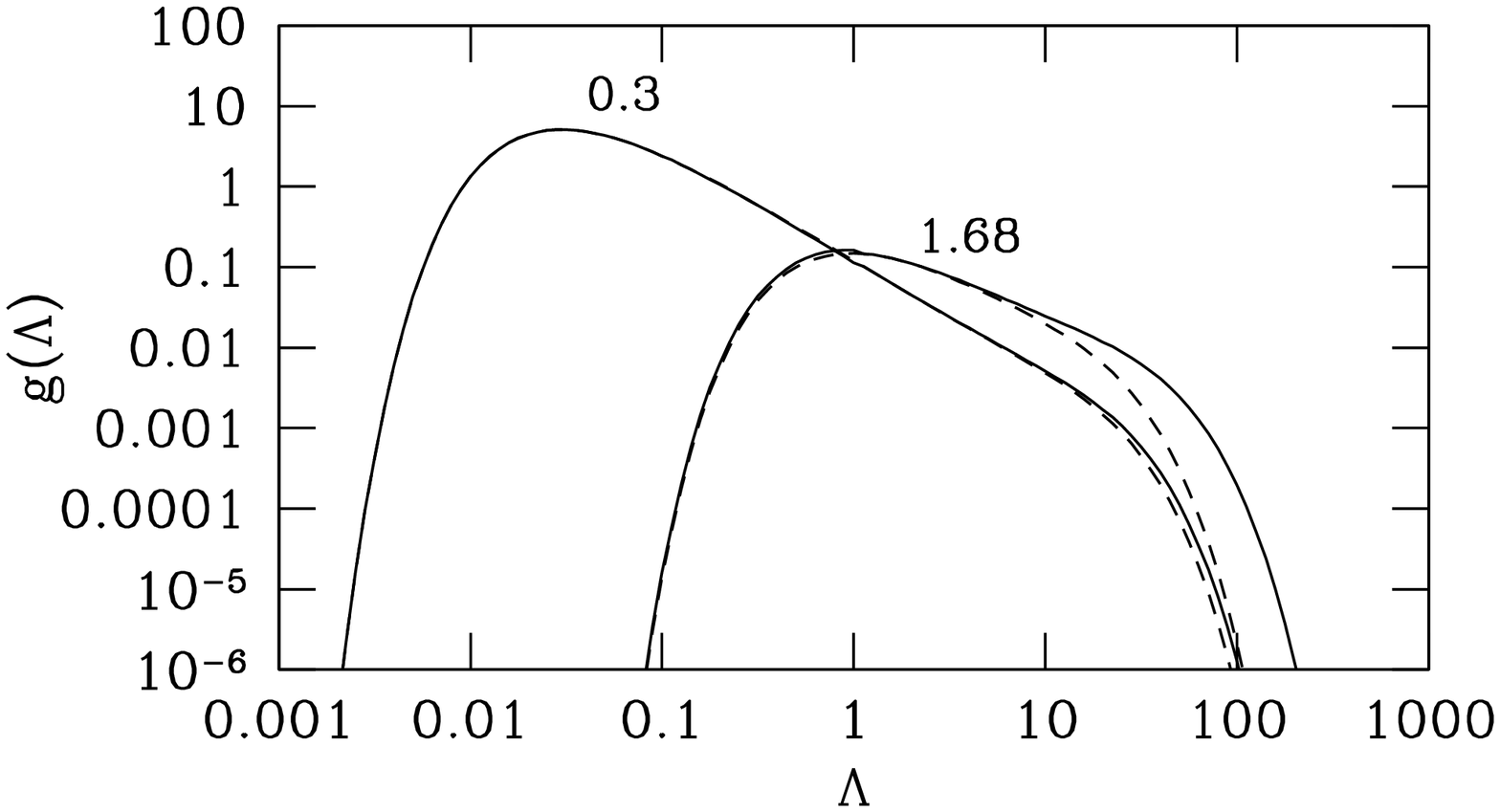,height=9.0cm,width=9.0cm}
\psfig{figure=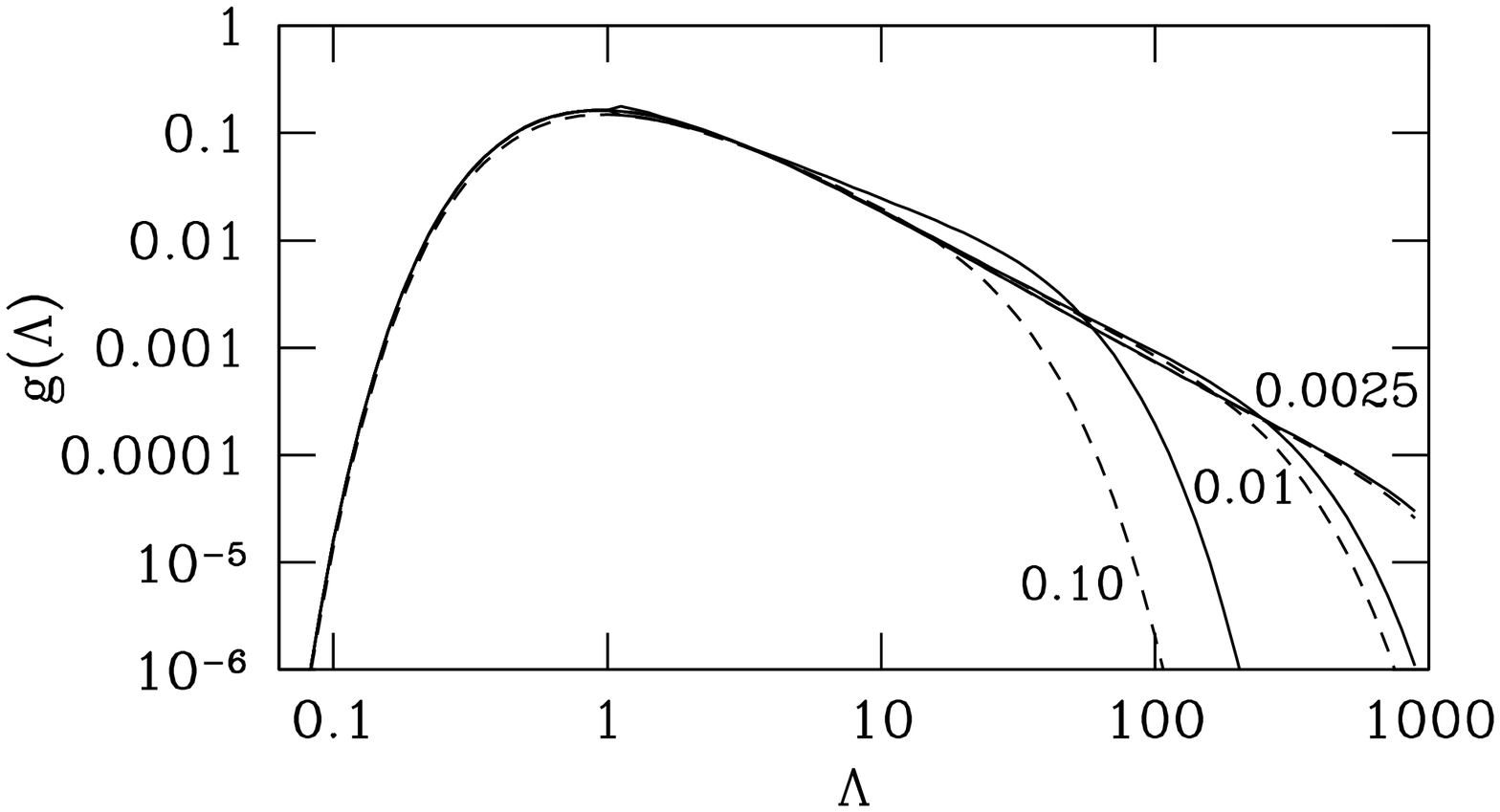,height=9.0cm,width=9.0cm}}
\vspace{-0.3cm}
\caption{\small Pseudo mass functions for order-1 processes obtained
with the Laplace transform method (continuous lines) and with the
mirror image method (dashed lines). Left ($A=0.1$): $f_{\rm
c}=0.3$ and 1.68. Right ($f_{\rm c}=1.68$): $A=0.10$, 0.01,
0.0025. Note the different scales of the two figures.}
\label{FIG_RIC}
\end{figure}

Since the absorbing barrier at $f_{\rm c}$ is constant, the Laplace
transform method (Sect.\,\ref{MASS_WIEN_1}) can be applied. The
Laplace transform of the corresponding Kolmogorov's backward equation
(e.g., Beekman 1975 and the references given therein) leads to the
pseudo mass function (definitions of the specific mathematical
functions used in the following can be found in, e.g., Abramowitz \&
Stegun 1979),
\begin{equation}\label{RIC1}
g^*(\lambda,f_{\rm c})\,=\,\exp{\left(-\frac{AS^2}{2D}\right)}\,
\frac{D_{-\lambda/A}(0)}{D_{-\lambda/A}(-f_{\rm c}\sqrt{\frac{2A}{D}})}\,=\,
\frac{\Phi_{D,A}(\lambda,0)}{\Phi_{D,A}(\lambda,f_{\rm c})}\,,
\end{equation}
where $D_\alpha(z)$ is the parabolic cylinder function with
$\alpha=-\Lambda/A$,
\begin{equation}\label{RIC2}
\Phi_{D,A}(\lambda,f_{\rm c})\,=\,\frac{1}{\Gamma(\frac{\lambda}{2A})\,\Gamma(\frac{1+\lambda/A}{2})}\,
\sum_{n=0}^{\infty}\,\frac{\left(2f_{\rm c}\sqrt{\frac{A}D}\right)^n}{n!}\,\Gamma\left(\frac{n+\lambda/A}{2}\right)
\end{equation}
an auxillary function, and $\Gamma(\cdot)$ the Euler gamma
function. The inversion of the Laplace transform appeared recently in
the mathematical literature (see Ricciardi \& Sato 1988) and involves
the zeros $\lambda=\lambda_\pi(f_{\rm c})$ of (\ref{RIC2}) -- they are
all negative -- giving the pseudo mass function
\begin{equation}\label{RIC3}
g(\Lambda,f_{\rm c})\,=\,\sum_{p=0}^{\infty}\,K_p(f_{\rm c})\,
\exp\left[{\lambda_p(f_{\rm c})\Lambda}\right]\,,
\end{equation}
where
\begin{equation}\label{RIC4}
K^{-1}_p(f_{\rm c})\,=\,\frac{1}{2A}\,\sum_{n=0}^\infty\,\frac{\left(2f_{\rm c}\sqrt{\frac{A}{D}}\right)^n}{n!}\,
\frac{\Gamma\left(\frac{n+\lambda_p/A}{2}\right)}{\Gamma\left(\frac{\lambda_p}{2A}\right)}\,
\left[\,\Psi\left(\frac{n+\lambda_p/A}{2}\right)\,-\,\Psi\left(\frac{1+\lambda_p/A}{2}\right)\,-\,
\Psi\left(\frac{\lambda_p}{2A}\right)\right]\,,
\end{equation}
and $\Psi(\cdot)$ being the digamma function. The asymptotic behaviour
of (\ref{RIC3}) may be summarized as
\begin{equation}\label{RIC5}
g(\Lambda,f_{\rm c})\,\sim\,\left\{
\begin{array}{r@{\quad:\quad}l}
\frac{f_{\rm c}}{\sqrt{2\pi D}} 
\left(\frac{1}{\Lambda}\right)^{\frac{3}{2}}\,e^{-f_{\rm c}^2/(2\Lambda)}
& \Lambda\,\, {\rm small} \\
K_0(f_{\rm c})\,e^{-|\lambda_0|\Lambda} &
\Lambda \,\,{\rm large} \end{array} \right.\,.
\end{equation}
The pseudo mass function can thus be expressed as the sum of
exponential functions with negative exponents ($\lambda_p<0$) and
reduces to single exponentials for small and for large $\Lambda$. For
$A\rightarrow 0$, Eq. (\ref{RIC3}) tends to the Press-Schechter
pseudo mass function. \\

\begin{figure}
\vspace{-2.7cm}
\centerline{\hspace{-9.0cm}
\psfig{figure=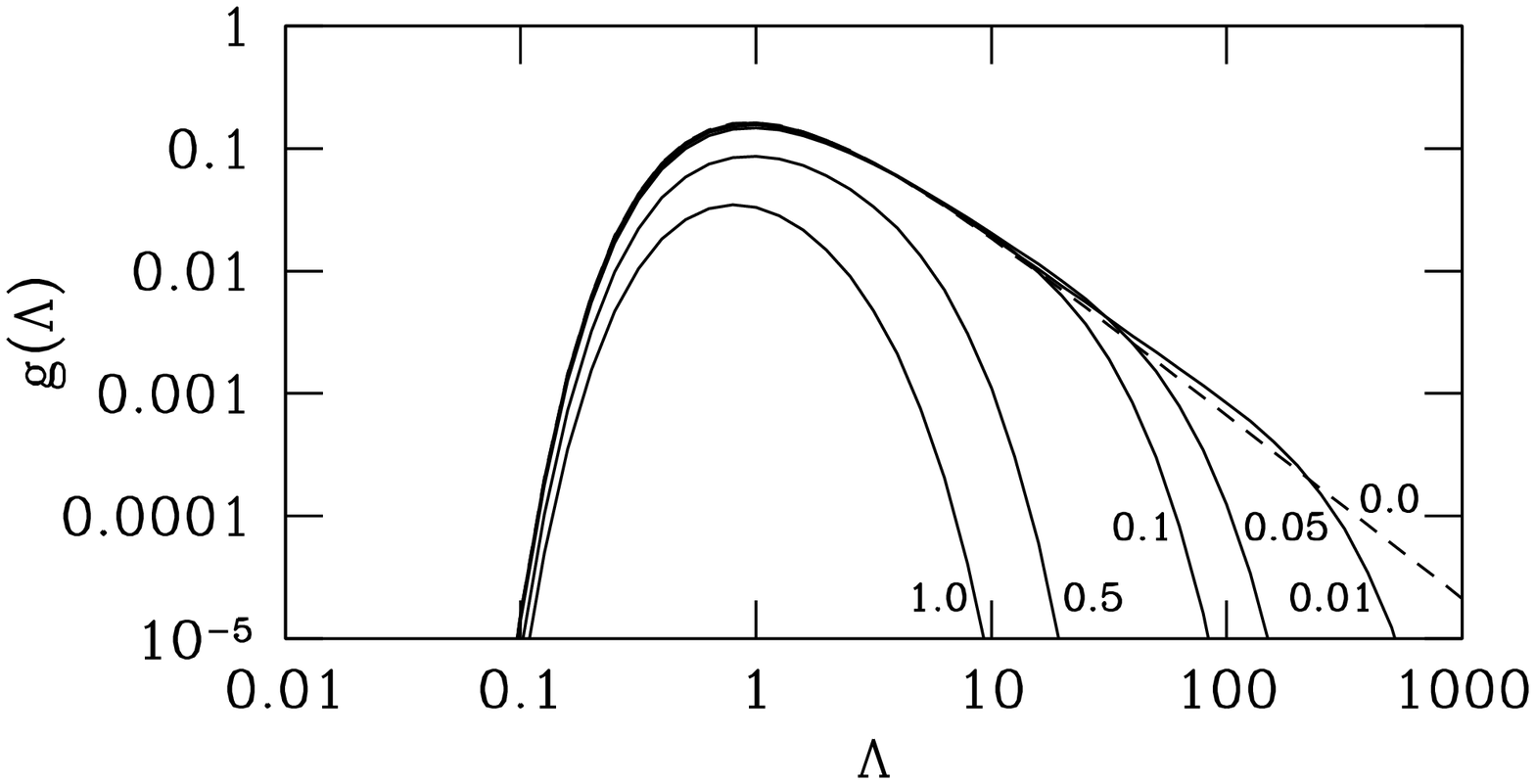,height=9.0cm,width=9.0cm}}
\vspace{-9.0cm}
\centerline{\hspace{+9.0cm}
\psfig{figure=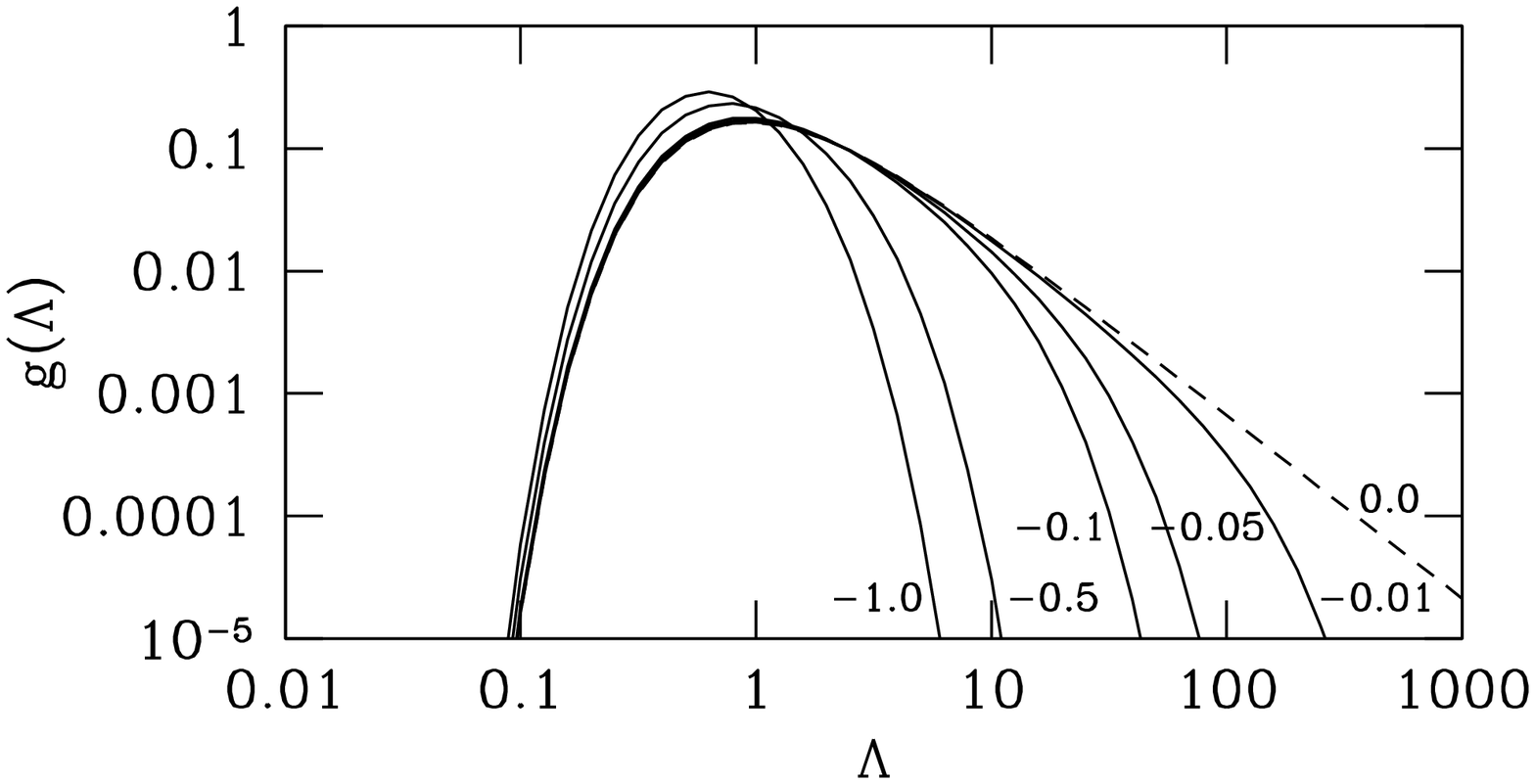,height=9.0cm,width=9.0cm}}
\vspace{-1.8cm}
\caption{\small Pseudo mass functions from order-1 processes obtained
with the mirror image method for different $A$ values (continuous
lines). For comparison, the Press-Schechter pseudo mass function
($A=0$) is given (dashed lines).}
\label{FIG_OU}
\end{figure}

These results are compared to the closed-form analytic approximation
obtained with the mirror image method (see
Sect.\,\ref{MASS_WIEN_1}). The unconstrained Fokker-Planck equation
has the solution
\begin{eqnarray}\label{ORNUHL1}
\Pi_1(\tilde{f}_0,\tilde{f},\Lambda)\,=\,\frac{1}{\sqrt{2\pi}}\,
\left[\frac{2A}{D(1-e^{-2A\Lambda})}\right]^{\frac{1}{2}}
\exp\left[-\frac{A(\tilde{f}-\tilde{f}_0e^{-A\Lambda})^2}
{D(1-e^{-2A\Lambda})}\right]\,,
\end{eqnarray}
where all sample paths start at $\tilde{f}_0$. The superposition
of two sources of sample paths gives the constrained solution
\begin{eqnarray}\label{ORNUHL2}
\Pi(\tilde{f},\Lambda)\,=\,\frac{1}{\sqrt{2\pi}}
\left[\frac{2A}{D(1-e^{-2A\Lambda})}\right]^{\frac{1}{2}}
\left[e^{-\frac{A\tilde{f}^2}{D(1-e^{-2A\Lambda})}}\,-
\,e^{\frac{4f_{\rm c}^2 Ae^{-A\Lambda}(e^{-A\Lambda}-1)}{D(1-e^{-2A\Lambda})}}\,
e^{-\frac{A(\tilde{f}-2f_{\rm c} e^{-A\Lambda})^2}{D(1-e^{-2A\Lambda})}}\right]\,.
\end{eqnarray}
Taking into account the boundary conditions (\ref{GWIEN4}) we obtain
\begin{eqnarray}\label{ORNUHL4}
g(\Lambda,f_{\rm c})\,=\,\frac{f_{\rm c}}{\sqrt{2\pi D}}\,
\left(\frac{2A}{1-e^{-2A\Lambda}}\right)^{\frac{3}{2}}
\,\exp\left[-\frac{Af_{\rm
c}^2}{D(1-e^{-2A\Lambda})}-A\Lambda\right]\,.
\end{eqnarray}
As for order-0 processes, $\Lambda$ is given by (\ref{WIN7}). Note
that for fixed $f_{\rm c}$ {\it all} pseudo mass functions agree
in the low-$\Lambda$ range, i.e., for large masses; differences are
seen in the high-$\Lambda$ range, i.e., for low masses. For small
$f_{\rm c}$ or small $A$ the solutions obtained with the Laplace
transform method can be approximated remarkebly well by the
closed-form analytic results obtained with the mirror image method
although the strict symmetry of the diffusion process is distorted by
the $\tilde{f}$-dependence of the drift coefficient.  For small
$A$ both solutions tend to the Press-Schechter pseudo mass function
(see Figs.\,\ref{FIG_RIC} and \ref{FIG_OU}).

\subsection{Mass functions}\label{MASS_OU_2}

Equating the process variance (\ref{OU2a}) with the mass variance,
$\sigma^2(R)$, and using (\ref{WIN7}) gives the relation which
determines the window function,
\begin{eqnarray}\label{WIN15}
\frac{1}{2\pi^2}\int_0^\infty dk\,k^2\,|W_R(k)|^2\,P(k)\,
=\,\frac{D}{2A}\left\{1-
\exp\left[-\frac{A}{\pi^2}\int_0^{\frac{1}{R}}dk\,k^2P(k)\right]\right\}\,.
\end{eqnarray}
For a given diffusion process the filter function and thus the region
where material ultimately collapses to form a virialized halo cannot
be chosen in any case independently from the power spectrum of the
Gaussian random field. 

For scale-invariant power spectra the filter function consistent with
the density field and diffusion process is
\begin{equation}\label{WIN16}
W_R(k)\,=\,\left\{ \begin{array}{r@{\quad:\quad}l}
\exp\left(-\frac{A\,P_0\,k^{n+3}}{2\pi^2(n+3)}\right) & k\leq 1/R \\ 
0 & {\rm else} \end{array} \right.\,. 
\end{equation}
This is a particular solution of the singular integral equation
(\ref{WIN15}) found `by eye'. Possible ambiguous solutions in the form
of strongly fluctuating window functions in $k$-space are physically
not very interesting and can thus be disgarded.  The normalization,
$D=1$, ensures $W_R(k=0)=1$. For $A>0$ it is an exponential filter
truncated at the wavenumber $1/R$. For large $R$
Eq. (\ref{WIN16}) resembles the sharp $k$-space filter whereas
for small $R$ the wings of the exponential get more
important. Consider first a power spectrum with $n=-2$ (the more
general case $n>-3$ is treated thereafter). For this realistic case
the truncated exponential filter (\ref{WIN16}) transformed into
configuration space is
\begin{eqnarray}
W_{R}(r)\,=\frac{e^{-\frac{AP_0}{2\pi^2R}}}
{2\pi^2r\left[\left(\frac{AP_0}{2\pi^2}\right)^2+r^2\right]}
\left\{-\left[\frac{AP_0}{2\pi^2R}+
\frac{\left(\frac{AP_0}{2\pi^2}\right)^2-r^2}
{\left(\frac{AP_0}{2\pi^2}\right)^2+r^2}\right]
\sin\left(\frac{r}{R}\right)\right.\,-\nonumber
\end{eqnarray}
\begin{equation}\label{WIN161}
\hspace{3.0cm}\left[\frac{r}{R}+
\frac{AP_0r}{\pi^2\left((\frac{AP_0}{2\pi^2})^2+r^2\right)}\right]
\cos\left(\frac{r}{R}\right) \left. +\,\frac{AP_0r\,e^{\frac{AP_0}{2\pi^2R}}}
{\pi^2\left[\left(\frac{AP_0}{2\pi^2}\right)^2+r^2\right]}\right\}\,.
\end{equation}
Eq. (\ref{WIN161}) corresponds to (\ref{WIN71}) as $A\rightarrow 0$
(Fig.\,\ref{FIG_FILT}). The cumulative mass and volume at the filter
scale $R$ are determined by the inverse of $W_R(r)$ as $r\rightarrow
0$, resulting in
\begin{equation}\label{WIN162}
M_{\rm E}\,=\,\bar{\rho}_0\,V_{\rm E}\,,\quad\quad V_{\rm
E}\,=\,2\pi^2\,\left(\frac{AP_0}{2\pi^2}\right)^2\,e^{\frac{AP_0}{2\pi^2R}}\,
\left[\frac{4\pi^2}{AP_0}\left(e^{\frac{AP_0}{2\pi^2R}}-1\right)\,-
\,\frac{2}{R}\,-\,\frac{AP_0}{2\pi^2R^2}\right]^{-1}\,,
\end{equation}
with the limits
\begin{equation}\label{WIN163}
V_{\rm E}\,=\,\left\{ \begin{array}{r@{\quad:\quad}l}
6\pi^2\,R^3 & R\gg\frac{AP_0}{2\pi^2}\\ 
\frac{\pi^2}{(2\pi^2)^3}\,(AP_0)^3 & R\ll\frac{AP_0}{2\pi^2} \end{array} \right.\,.
\end{equation}
The $M(\Lambda)$ relation thus becomes
\begin{equation}\label{WIN164}
M_{\rm
E}\,=\,\bar{\rho}_0\,\left(\frac{AP_0}{2\pi^2}\right)^2\,P_0\,e^{A\Lambda}\,
\left[\frac{2}{A}\left(e^{A\Lambda}-1\right)\,-\,2\Lambda\,-A\Lambda^2\right]^{-1}\,,
\end{equation}
which can be solved numerically for $\Lambda$ to compute the mass
function (Eq. \ref{GWIEN91}). Ratios of the mass functions induced by
order-1 processes with positive drift parameters to the
Press-Schechter mass function are shown in
Fig.\,\ref{FIG_NM_W_OU}. Eq. (\ref{WIN164}) can be approximated by
\begin{equation}\label{WIN165}
M_{\rm
E}\,\approx\,\frac{3\bar{\rho}_0P_0^3}{4\pi^4}
\left(\frac{1}{\Lambda^3}\,+\,\frac{A^3}{6}\right)\,,
\end{equation}
leading to the $\Lambda(M)$ relation (we omit the subscript of $M$)
\begin{equation}\label{WIN166a}
\Lambda\,\approx\,\left(\frac{4\pi^4M}{3\bar{\rho}_0P_0^3}-\frac{A^3}{6}\right)^{-\frac{1}{3}}\,.
\end{equation}
and with the approximation (\ref{ORNUHL4}) to the mass function
\begin{equation}\label{WIN166}
n(M)\,\approx\,\frac{4\pi^4}{9\sqrt{2\pi}}\,\frac{f_{\rm
c}}{MP_0^3}
\,\Lambda^4\,\left(\frac{2\,A}{1-e^{-2A\Lambda}}\right)^{\frac{3}{2}}\,
\exp\left(-\frac{A\,f_{\rm c}^2}
{1-e^{-2A\Lambda}}\,-\,A\Lambda\right)\,.
\end{equation}
It is seen that the mass function gives more objects with low mass and
similar numbers of objects with high mass compared to the
Press-Schechter case (see Fig.\,\ref{FIG_NM_W_OU}). The differences
increase with increasing drift parameter and illustrate the effects of
the different shapes of the filter functions (see
Fig.\,\ref{FIG_FILT}) on the mass function.

For scale-invariant power spectra with index $n>-3$, expansion of
(\ref{WIN16}) and pice-wise integration gives the general $M(\Lambda)$
relation,
\begin{equation}\label{WIN20}
M_{\rm E}(\Lambda)\,=\,\bar{\rho}_0\,V_{\rm E}\,,\quad
V^{-1}_{\rm E}\,=\,W_R(r=0)\,=\,\frac{1}{2\pi^2}\,\int_0^\infty\,dk\,k^2\,W_R(k)\,=\,
\frac{\sum\limits_{m=0}^\infty\frac{(-1)^m\,A^m}{m!\,
\left[m\left(\frac{n+3}{3}\right)+1\right]}\,\Lambda^m}
{6\pi^2\,\left[\frac{P_0\Lambda}{2\pi^2(n+3)}\right]^{\frac{3}{n+3}}}\,,
\end{equation}
from which the mass function can be obtained numerically in the same
way as from (\ref{WIN164}). For $A\rightarrow 0$ only the $m=0$ term
in the numerator of (\ref{WIN20}) gives a contribution to the sum,
thus reproducing the volume of the sharp $k$ space filter.

\section{Summary and discussion}\label{DISCUSS}

The present paper concentrates on the development of methods to derive
analytic mass functions for given stochastic diffusion processes. For
practical purposes the analysis is restricted to order-0 and order-1
processes. Whereas order-0 processes are characterized by drift
coefficients which are independent of the filtered density contrast,
order-1 processes have a linear dependence on the filtered density
contrast. It is seen that in order-1 stochastic diffusion processes,
although derived within the framework of Markovian processes,
realistic nonzero covariances appear in a very natural way without
sampling over the diffusion (merging) history. The correlation between
two mass scales gets important especially for small masses. For
realistic cases the filter functions corresponding to the diffusion
processes are derived. Whereas for all order-0 processes the filter
function is the sharp $k$-space filter, for order-1 processes
exponential filters are found. The effects of non-sharp $k$-space
filters on the mass function are in the low-mass range where an excess
compared to the standard Press-Schechter expectation is found. Future
investigations should work with higher order diffusion processes in
order to get more realistic potential wells (determined by the drift
term) for sample diffusion. Of special interest are asymmetric
potentials which might give a more appropriate handling of the
asymmetric distribution of $\tilde{f}$ within $[-1,\infty)$.

Nonetheless, do we really need a discussion of more general diffusion
processes?  From the practical point of view one might argue that the
Press-Schechter mass function, although derived under very specific
assumptions, is well confirmed by comparison with N-body simulations
(e.g., Efstathiou et al.  1988, White, Efstathiou \& Frenk 1993, Lacey
\& Cole 1994, Mo et al. 1996) and should thus suffice to be used as a
simple fitting formula. The mass resolution and dynamic range of the
simulations are, however, too low to test the prescription in
detail. New results show indeed significant deviations from the
analytical expectation (e.g., Jing 1998, Sheth \& Tormen 1999,
Governato et al. 1999). In the light of the discussion given above one
might expect such deviations from the Press-Schechter
formula. Consequently one should avoid any unnecessary conditions on
the type of diffusion process. The new coefficients offer a more
flexible summary of the new mass functions.

In forthcoming papers we want to derive consistent mass and filter
functions for more complicated stochastic diffusion processes and
compare the results with detailed N-body simulations and observed mass
and luminosity functions derived from physically well-defined X-ray
selected samples of clusters of galaxies.

\begin{acknowledgements}                                                        
We would like to thank J\"org Retzlaff for a critical reading of the
manuscript and the anonymous referee for many useful
comments. PS. thanks for support by the Verbundforschung under the
grant No.\,50\,OR\,9708\,35.
\end{acknowledgements}

\begin{appendix}

\section{Order-1 stochastic diffusion processes}\label{APP_OU}

The It\^o stochastic differential equation of these processes is
\begin{equation}\label{OUA1}
d\tilde{f}(\Lambda)\,=\,-A\,\tilde{f}\,d\Lambda\,+\,\sqrt{D}\,dw(\Lambda)\,.
\end{equation}
Early versions of this equation can be found in Uhlenbeck \& Ornstein
(1930) and Doob (1942). Eq. (\ref{OUA1}) can be solved by introducing
the variable $\tilde{u}=\tilde{f}e^{A\Lambda}$, and using the It\^o
formula for a sufficiently smooth function, $f$, of
$\tilde{f}(\Lambda)$ satisfying (\ref{ITO}),
\begin{equation}\label{ITOFORM}
df[\tilde{f}(\Lambda)]\,=\, \left[\frac{\partial
f}{\partial\Lambda}\,-\,A_1(\tilde{f}(\Lambda),\Lambda)\frac{\partial
f}{\partial\tilde{f}}\,+\,
\frac{1}{2}A_2(\tilde{f}(\Lambda),\Lambda)
\frac{\partial^2f}{\partial\tilde{f}^2}\right]\,d\Lambda\,+\,
\sqrt{A_2(\tilde{f}(\Lambda),\Lambda)}\frac{\partial
f}{\partial\tilde{f}}dw(\Lambda)\,,
\end{equation}
where the partial derivatives are evaluated at
$(\Lambda,\tilde{f}(\Lambda))$. For $f$ linear in
$\tilde{f}$ we have $\partial^2f/\partial\tilde{f}^2=0$, and
(\ref{ITOFORM}) reduces to the usual chain rule. If $f=\tilde{u}$ this
gives $d\tilde{u}=\sqrt{D}\,e^{A\Lambda}\,dw(\Lambda)$, which can be
integrated and yields for the initial condition, $\tilde{f}(0)=0$,
\begin{equation}\label{OUA3}
\tilde{f}(\Lambda)\,=\,\sqrt{D}\,\int_0^{\Lambda}e^{-A(\Lambda-\Lambda')}\,dw(\Lambda')\,.
\end{equation}
The increments, $dw(\Lambda)$, of the fundamental (Wiener) process
have a Gaussian distribution (Sect.\,\ref{GENERAL}). Hence, the linear
superposition in (\ref{OUA3}) shows that the $\tilde{f}$ are
again Gaussian distributed with zero mean, the covariance
\begin{equation}\label{OUA4}
{\rm E}[\tilde{f}(\Lambda_1)\,\tilde{f}(\Lambda_2)\,]\,=
\,D\,\cdot\,{\rm
E}\left[\int_0^{\Lambda_1}e^{-A(\Lambda_1-\Lambda_1')}dw(\Lambda_1')
\int_0^{\Lambda_2}e^{-A(\Lambda_2-\Lambda_2')}dw(\Lambda_2')\right]\,=\,
\frac{D}{2A}\,\left[\,e^{-A|\Lambda_1-\Lambda_2|}\,-\,
e^{-A(\Lambda_1+\Lambda_2)}\right]\,,
\end{equation}
and the instantaneous variance
\begin{equation}\label{OUA5}
{\rm var}\left[\tilde{f}(\Lambda)\right]\,=\,\frac{D}{2A}
\left(1\,-\,e^{-2A\Lambda}\right)\,.
\end{equation}
Nonzero covariances between $\tilde{f}(\Lambda)$ and the increments
$\Delta\tilde{f}(\Lambda+f\Lambda)$ lead to more realistic
correlations between different parts of each sample path, and thus to
more realistic shapes of the corresponding process filters. For $A>0$,
and $\Lambda_1,\Lambda_2\gg 1$, the last term in (\ref{OUA4}) can be
neglected, and the process has, as expected, a stationary covariance
function which decreases exponentially with increasing distance
$|\Lambda_1-\Lambda_2|$, corresponding to a non-white power
spectrum. For $A<0$ less fast decreasing covariances are found.

The process can be generalized to the case with $\Lambda$-dependent
coefficients,
\begin{equation}\label{OUA6}
d\tilde{f}(\Lambda)\,=\,-A(\Lambda)\,\tilde{f}\,d\Lambda\,+\,
\sqrt{D(\Lambda)}\,dw(\Lambda)\,.
\end{equation}
For $\tilde{f}(\Lambda=0)\,=\,0$ the
solution of (\ref{OUA6}) is
\begin{equation}\label{OUA7}
\tilde{f}(\Lambda)\,=\,\int_0^{\Lambda}dw(\Lambda')\,
\sqrt{D(\Lambda')}\,\exp\left[-\int_{\Lambda'}^{\Lambda}A(s)ds\right]\,,
\end{equation}
giving the covariance
\begin{eqnarray}\label{OUA9}
{\rm
E}[\tilde{f}(\Lambda_1)\,\tilde{f}(\Lambda_2)]\,=\,
\int_0^{{\rm
min}(\Lambda_1,\Lambda_2)}d\Lambda'\,D(\Lambda')
\quad\exp\left[\,-\int_{\Lambda'}^{\Lambda_1}\,A(s)\,ds\,-\,
\int_{\Lambda'}^{\Lambda_2}\,A(s')\,ds'\right]\,,
\end{eqnarray}
and the instantaneous variance
\begin{equation}\label{OUA8}
{\rm
var}[\tilde{f}(\Lambda)]\,=\,\int_0^{\Lambda}d\Lambda'\,D(\Lambda')\,
\exp\left[-\int_{\Lambda'}^{\Lambda}2\,A(s)ds\right]\,.
\end{equation}

\end{appendix}


\begin{thebibliography}{}

\bibitem{} Abramowitz, M., Stegun, I.A., 1979, Handbook of
mathematical functions (Dover, New York)

\bibitem{} Arnold, L., 1974, Stochastic Differential Equations: Theory 
and Applications  (John Wiley and Sons, New York)

\bibitem{} Beekman, J.A., 1975, J.\,App.\,Prob., 12, 107

\bibitem{} Bernardeau, F., 1994, ApJ, 427, 51

\bibitem{} Bertschinger, E., 1985, ApJS, 58, 39

\bibitem{} Binggeli, B., Sandage, A.R., Tammann, G.A., 1988, ARAA, 26,
509

\bibitem{} B\"ohringer, H., et al., 1998, The Messenger, 94, 21

\bibitem{} Bond, J.R., Cole, S., Efstathiou, G., Kaiser, N., 1991,
ApJ, 379, 440

\bibitem{} Borgani, S., Rosati, P., Tozzi, P., Norman, C., 1999, ApJ,
517, 40

\bibitem{} Bower, R.J., 1991, MNRAS, 248, 332

\bibitem{} Cavaliere, A., Menci, N., 1994, ApJ, 435, 528

\bibitem{} Chiu, W.A., Ostriker, J.P.,   Strauss, M., 1998, ApJ, 494,
479

\bibitem{} Cole, S., Kaiser, N., 1988, MNRAS, 233, 637

\bibitem{} Cole, S., Kaiser, N., 1989, MNRAS, 237, 1127

\bibitem{} Cox, D.R., Miller, H.D., 1965, The Theory of Stochastic
Processes (Chapmann Hall, London)

\bibitem{} Doob, J.L., 1942, Ann. Math., 43, 351

\bibitem{} Efstathiou, G., Frenk, C.S., White, S.D.M., and Davis, M.,
1988, MNRAS, 235, 715

\bibitem{} Eke, V.R., Cole, S., Frenk, C.S., 1996, MNRAS, 282, 263

\bibitem{} Ellis, R.S., Colless, M., Broadhurst, T., Heyl, J.,
Glazebrook, K., 1996, MNRAS, 280, 235

\bibitem{} Friedman, A., 1975, Stochastic Differential Equations and
Applications, Vol.\,1 (Academic Press, New York)

\bibitem{} Gardiner, C.W., 1997, Handbook of Stochastic Methods
(Springer Verlag, Berlin)

\bibitem{}Governato, F., Babul, A., Quinn, T., Tozzi, P., Baugh, C.M.,
Katz, N., Lake, G., 1999, MNRAS, 307, 949

\bibitem{} Gunn, J.E., Gott, J.R.\,III., 1972, ApJ, 176, 1

\bibitem{} Guzzo, L., et al., 1999, The Messenger, 95, 27

\bibitem{} Henry, J.P., Arnaud, K.A., 1991, ApJ, 372, 410

\bibitem{} It\^o, K.,   McKean Jr., H.P., 1974, Diffusion Processes
and their Sample Paths (Springer-Verlag, Berlin), 21

\bibitem{} Jedamzik, K., 1995, ApJ, 448, 1

\bibitem{} Jing, Y.P., 1998, ApJ, 503, L9

\bibitem{} Kauffmann, G., White, S.D.M., 1993, MNRAS, 261, 921

\bibitem{} Kitayama, T., Suto, Y., 1996, ApJ, 469, 480

\bibitem{} Kloeden, P.E., Platen, E., 1995, Numerical Solution of
Stochastic Differential Equations (Springer Verlag, Berlin)

\bibitem{} Lacey, C., Cole, S., 1993, MNRAS, 262, 627

\bibitem{} Lacey, C., Cole, S., 1994, MNRAS, 271, 676

\bibitem{} Landau, L.D., Lifschitz, E.M., 1966, Hydrodynamik (Akademie
Verlag, Berlin)

\bibitem{} Lee, J., Shandarin, S.F., 1998, ApJ, 500, 14

\bibitem{} Lilje, P.B., 1992, ApJ, 386, L33

\bibitem{} Lilly, S.J., Le\,Fevre, O., Crampton, D., Hammer, F.,
Tresse, L., 1995, ApJ, 455, 50

\bibitem{} Lucchin, F., Matarrese, S., 1988, ApJ, 330, 535

\bibitem{} Matarrese, S., Coles, P., Lucchin, F., Moscardini, L.,
1997, MNRAS, 286, 115

\bibitem{} Mathiesen, B., Evrard, A.E., 1998, MNRAS, 295, 769

\bibitem{} Mo, H.J.  White, S.D.M., 1996, MNRAS, 282, 347

\bibitem{} Mo, H.J., Jing, Y.P., White, S.D.M., 1996, MNRAS, 282, 1096

\bibitem{} Monaco, P., 1995, ApJ, 447, 23

\bibitem{} Narayan, R., White, S.D.M., 1988, MNRAS, 231, 97p

\bibitem{} Navarro, J.F., Frenk, C.S., White, S.D.M., 1996, ApJ, 462,
563

\bibitem{} Peebles, P.J.E., 1990, ApJ, 365, 27

\bibitem{} Peacock, J.A., 1999, Cosmological Physics (Cambridge
Univ. Press, Cambridge)

\bibitem{} Peacock, J.A., Heavens, A.F., 1990, MNRAS, 243, 133

\bibitem{} Press, W.H., Schechter, P., 1974, ApJ, 187, 425

\bibitem{} Ricciardi, L.M., Sato, S., 1988, J.\,App.\,Prob., 25, 43

\bibitem{} Risken, H., 1984, The Fokker-Planck Equation
(Springer-Verlag, Berlin)

\bibitem{} Sheth, R.K., Tormen, G., 1999, MNRAS, 308, 119

\bibitem{} Uhlenbeck, G.E., Ornstein, L.S., 1930, Phys. Rev., 36, 823

\bibitem{} Valageas, P., Schaeffer, R., 1997, A A, 328, 435

\bibitem{} Wax, N., 1954, Selected papers on noise and stochastic
processes (Dover Publ. Inc., New York)

\bibitem{} White, S.D.M., 1996, in Cosmology and Large-Scale
Structure, eds. R. Schaeffer, J. Silk, M. Spiro and J. Zinn-Justin
(Elsevier, Dordrecht), 349

\bibitem{} White, S.D.M., 1997, in The Evolution of the Universe,
eds. G. B\"orner and S. Gottl\"ober (Wiley   Sons, Chicester), 227

\bibitem{} White, S.D.M., Frenk, C.S., 1991, ApJ, 379, 25

\bibitem{} White, S.D.M., Efstathiou, G., Frenk, C.S., 1993, MNRAS,
262, 1023

\bibitem{} Wiener, N., 1930, Acta Math., 55, 117

\bibitem{} Yano, T., Nagashima, M., Gouda, N., 1996, ApJ, 466, 1

\end{thebibliography}
\end{document}